\documentclass[10pt]{article}
\usepackage{amsmath,amsfonts,amssymb,amsthm}
\usepackage{graphicx}
\usepackage{float}
\usepackage[latin1]{inputenc}
\usepackage{hyperref}
\usepackage{dsfont}
\usepackage{subcaption}
\usepackage{color}
\usepackage{multirow}  

\newcommand{\ket}[1]{\left|#1\right\rangle}
\newcommand{\bra}[1]{\left\langle #1\right|}
\newcommand{\ra}{\rangle}

\newcommand{\nn}{\nonumber}

\newcommand{\Wthree}[6]{\left(\begin{array}{ccc} #1 & #2 & #3 \\ #4 & #5 & #6 \end{array}\right)}
\newcommand{\Wfour}[9]{\left(\begin{array}{cccc} #1 & #2 & #3 & #4 \\ #5 & #6 & #7 & #8 \end{array}\right)^{(#9)}}
\newcommand{\Wsix}[6]{\left \{ \begin{array}{ccc} #1 & #2 & #3 \\ #4 & #5 & #6 \end{array}\right \} }
\newcommand{\Wnine}[9]{\left \{ \begin{array}{ccc} #1 & #2 & #3  \\ #4 & #5 & #6 \\ #7 & #8 & #9 \end{array}\right \} }

\newcommand{\coolname}{\texttt{sl2cfoam} }

\begin{document}

\title{Numerical methods for EPRL spin foam transition amplitudes and Lorentzian recoupling theory}

\author{\Large{Pietro Don\`a${}^a$\footnote{pxd81@psu.edu}, Giorgio Sarno${}^b$ \footnote{sarno@cpt.univ-mrs.fr}}
\smallskip \\ 
\small{$^a$ Institute for Gravitation and the Cosmos \& Physics Department,}\\
\small{Penn State, University Park, PA 16802, USA} \\
\small{$^b$ Aix Marseille Univ, Universit\'e de Toulon, CNRS, CPT, Marseille, France}
}

\date{\today}

\maketitle

%----------------------------------------------------------------------------
\begin{abstract}
\noindent
The intricated combinatorial structure and the non-compactness of the Lorentz group have always made the computation of $SL(2,\mathbb{C})$ EPRL spin foam transition amplitudes a very hard and resource demanding task. With \coolname we provide a C-coded library for the evaluation of the Lorentzian EPRL vertex amplitude.
We provide a tool to compute the Lorentzian EPRL 4-simplex vertex amplitude in the intertwiner basis and some utilities to evaluate SU(2) invariants, booster functions and $SL(2,\mathbb{C})$ Clebsch-Gordan coefficients. We discuss the data storage, parallelizations, time, and memory performances and possible future developments.  
\end{abstract}
%----------------------------------------------------------------------------

\tableofcontents

%----------------------------------------------------------------------------
\section{Introduction}
The spin foam framework provides an approach to the dynamics of loop quantum gravity that is both background independent and Lorentz covariant. The most successful model is the one proposed by Engle, Pereira, Rovelli and Livine (EPRL) \cite{Engle:2007wy,Livine:2007vk,Livine:2007ya} and independently by Freidel and Krasnov (-FK) \cite{Freidel:2007py}. The model is formulated as a sum over simplicial triangulations with given boundary providing transition amplitudes for (projected) spin network states of the canonical theory restricted to four valent nodes. This procedure is equivalent to a sum over histories of quantum geometries providing in this way a regularised version of the quantum gravity path integral.

A generalization to arbitrary complexes and spin-networks has been developed \cite{Kaminski:2009fm}. Nevertheless, in this work, we focus primarily on the simplicial formulation, and we provide all the tools for more general computations. 

Computations in this framework are particularly complex. Two of the most successful analytical calculations are: On the one hand, the presence of the discrete Regge action in the spin foam amplitude of a semiclassical state \cite{Barrett:2009gg,Barrett:2011xa}. Regge calculus provides a discretization of general relativity based on a triangulation. \footnote{The variables of Regge calculus are the simplex lengths (encoding the intrinsic curvature). The extrinsic curvature is encoded in the dihedral angles hinging on the triangles of the triangulation.} On the other hand, the graviton n-point function reproduces at the leading order the one of Regge calculus in the semi-classical limit \cite{Bianchi:2006uf,Speziale:2008uw,Bianchi:2009ri,Bianchi:2011hp}. This result is of particular historical importance since it revealed the shortcomings of the Barrett-Crane model  and led to the development of the EPRL-FK model. In particular, it has been shown that within the Barrett-Crane model it was not possible to reproduce the correct long-distance scaling of the graviton two-point function \cite{Alesci:2007tx,Alesci:2007tg,Alesci:2008ff}. Therefore, the model is unable to implement the simplicity constraints correctly. These are responsible for reducing the topological BF theory, the starting point of any spin foam model, to gravity.

Evaluating spin foam transition amplitudes for assigned boundary data is an arduous task. All the extraordinary results we mentioned are asymptotic computations performed in the large spin limit. At the present time, numerical calculations within the theory are very rare. Spin correlations have been computed with the old Barrett-Crane model \cite{Christensen:2007rv,Christensen:2009bi} or with the Lorentzian EPRL model but restricted to simple generalized spin foams \cite{Sarno:2018ses}. 

In this work, we introduce the library \coolname where we provide the tools to evaluate the EPRL spin foam vertex amplitude numerically. It is based on the decomposition of the EPRL vertex amplitude in terms of booster functions introduced in \cite{Speziale:2016axj}. These utilities are indicated for the study of vertex amplitudes with coherent boundary data and non, opening new windows on the covariant formulation of loop quantum gravity.

Some preliminary results have already been achieved. Using a primitive version of this code, it was possible to verify, for the first time, the asymptotic behavior of the topological BF SU(2) spin foam amplitude \cite{Dona:2017dvf}, but also to find the semiclassical region at relatively small values of the spins.
The transition amplitudes and spin correlations between two dipole spin networks, in the Lorentzian EPRL model, with up to two (non-simplicial) vertices has been studied numerically in \cite{Sarno:2018ses}. Even if restricted to simple vertices this was the first full evaluation of an EPRL transition amplitude. 
The numerical computation of the non-isotropic scaling of the booster functions, performed with a piece of this software, is a central ingredient in the recent estimation of the degree of divergence of the EPRL model \cite{Dona:2018pxq}. 

Moreover, we are finishing the numerical exploration of the Euclidean sector of the asymptotic of the Lorentzian EPRL vertex amplitude, reproducing with great accuracy the analytic asymptotic formula \cite{citaNoiEprl}.

With the access to a reliable and systematic way to perform numerical computation within the Lorentzian spin foam framework, in particular with more than a single vertex (or equivalently with many simplices), we are opening a window on a plethora of open questions. One question we plan to address soon is the so-called \textit{flatness problem}, firstly mentioned by Freidel and Conrady \cite{Conrady:2008mk}, later by Bonzom \cite{Bonzom:2009hw} and Hellmann and Kaminski \cite{Hellmann:2013gva}. They argue that the classical geometries reproduced by the EPRL model in the semiclassical limit are restricted to be flat and, as a result, the model itself is put in question, since GR in four dimensions admits curved spacetime solutions. A preliminary answer, in the Euclidean setting, has been given \cite{Oliveira:2017osu, Bayle:2016doe}, showing that curved geometric configurations are asymptotically allowed. We think that a numerical exploration is viable in Lorentzian setting and could help us shedding light on this issue.

The realm of applications of \coolname is not limited to technical results. The computation of physical observables requires the ability to evaluate spin foam transition amplitude with multiple vertices. The study of the black hole to white hole tunneling processes \cite{Christodoulou:2016vny, Christodoulou:2018ryl} and quantum cosmological models \cite{Bianchi:2010zs,Vidotto:2011qa,Sarno:2018ses} is the beginning of a long term project that aims to connect the quantum theory of gravity with observation. We hope that the release of the tools in \coolname will help these computations to flourish.

The purpose of the present work goes beyond giving the tools for the computation of Lorentzian EPRL transition amplitudes. In particular, we believe that the techniques we introduced and the algorithms we implemented can be used by the loop quantum gravity community on problems outside the spin foam context. For example, the numerical evaluation of $SL(2,\mathbb{C})$ Clebsch Gordan coefficients is a challenging and interesting mathematical problem on its own and the very efficient way we introduced to deal with SU(2) invariant can play a key role in exploring numerically the action of the hamiltonian operator in the canonical approach \cite{Alesci:2013kpa}.

We assume that the reader is familiar with the EPRL model, and refer to the original literature \cite{Engle:2007wy, Livine:2007vk, Livine:2007ya, Freidel:2007py} and existing reviews (e.g. \cite{Perez:2012wv,Carlo:2014}) for motivations, details and its relation to loop quantum gravity. We give a short review of the transition amplitude in Section \ref{sec:general}. 

In Section \ref{sec:library} we present in details how the library \coolname is structured. In Section \ref{sec:strategy} we give a general overview of the problems we faced and the solutions we propose, while in Sections \ref{sec:su2}, \ref{sec:booster} and \ref{sec:amplitude} we discuss its three main components, respectively the evaluation of SU(2) invariants, booster functions and finally the complete vertex amplitude. 

In the last two sections we present two additional utilities that computes the Livine-Speziale coherent states in Section \ref{sec:coherent} and the $SL(2,\mathbb{C})$ Clebsch-Gordan coefficients in Section \ref{sec:sl2c}.

All the tests shown in this paper have been performed on Intel(R) Xeon(R) E5-2687W v2 @ 3.40GHz CPUs provided by CPT servers in Marseille. If not stated otherwise the tests are performed on a single core.

The library \coolname can be found at: \url{https://bitbucket.org/giorgiosarno/sl2cfoam-1.0/}. We provide a \texttt{README} file in which we explain how to use the code. We include the test programs  we used to perform the benchmarks in this paper and can be seen as examples. The computation of a vertex amplitude for given data has to be computed only once. We include a compressed table of pre-calculated symbols, building blocks, and full amplitudes that we already computed in the repository. We will continuously update it in the future.

\section{The EPRL spin foam transition amplitude}
\label{sec:general}
The spin foam formalism introduces a partition function of a given closed 2-complex $\mathcal{C}$ as a state sum over $SU(2)$ spins $j_f$ on the faces and intertwiners $i_e$ on the edges: 
\begin{equation}
\label{eq:partitionf}
Z_{\mathcal{C}} = \sum_{j_f, i_e}  \prod_f A_f(j_f) \prod_e A_e(i_e) \prod_v A_v \left(j_f, \  i_e\right) \ .
\end{equation}
defined in terms of the face amplitude $A_f$, and the edge amplitude $A_e$ and the vertex amplitude $A_v$. Requiring the correct convolution property of the path integral at fixed boundary, the form of the face amplitude $A_f(j_f) := 2 j_f +1$ and the edge amplitude $A_e(i_e)=2i_e+1$ are fixed \cite{Bianchi:2010fj}. In this paper, we will focus on the numerical evaluation of the vertex amplitude $A_v$ defined by the EPRL model.

The vertex amplitude is built from the topological $SL(2,\mathbb{C})$ BF spin foam vertex amplitude by imposing, weakly, the simplicity constraints resulting into a restriction of the unitary irreducible representations in the principal series \cite{Engle:2007wy,Engle:2007uq}. To evaluate it in its original form one should perform four group integrals (one of the original five is redundant and has to be removed to guaratee finiteness \cite{Engle:2008ev}). Each group integral is, in general, a six dimensional unbounded highly oscillating integral for which numerical integration methods are very inefficient. To solve this issue, a more computational friendly form for the amplitude, has been proposed \cite{Speziale:2016axj}:
\begin{align}
\label{eq:vertexEPRL} 
A_v \left(j_f, \, i_e\right) &= \sum_{l_{f}, k_{e}} \left(\prod_{e} \left(2k_{e}+1\right) B_{4}(j_{f},l_{f};i_{e}, k_{e})\right) \{15j\}_{j_f,i_4}(l_{f}, k_{e}) \\ \nn
& = \sum_{l_{f}, k_{e}} \raisebox{-20mm}{ \includegraphics[width=7cm]{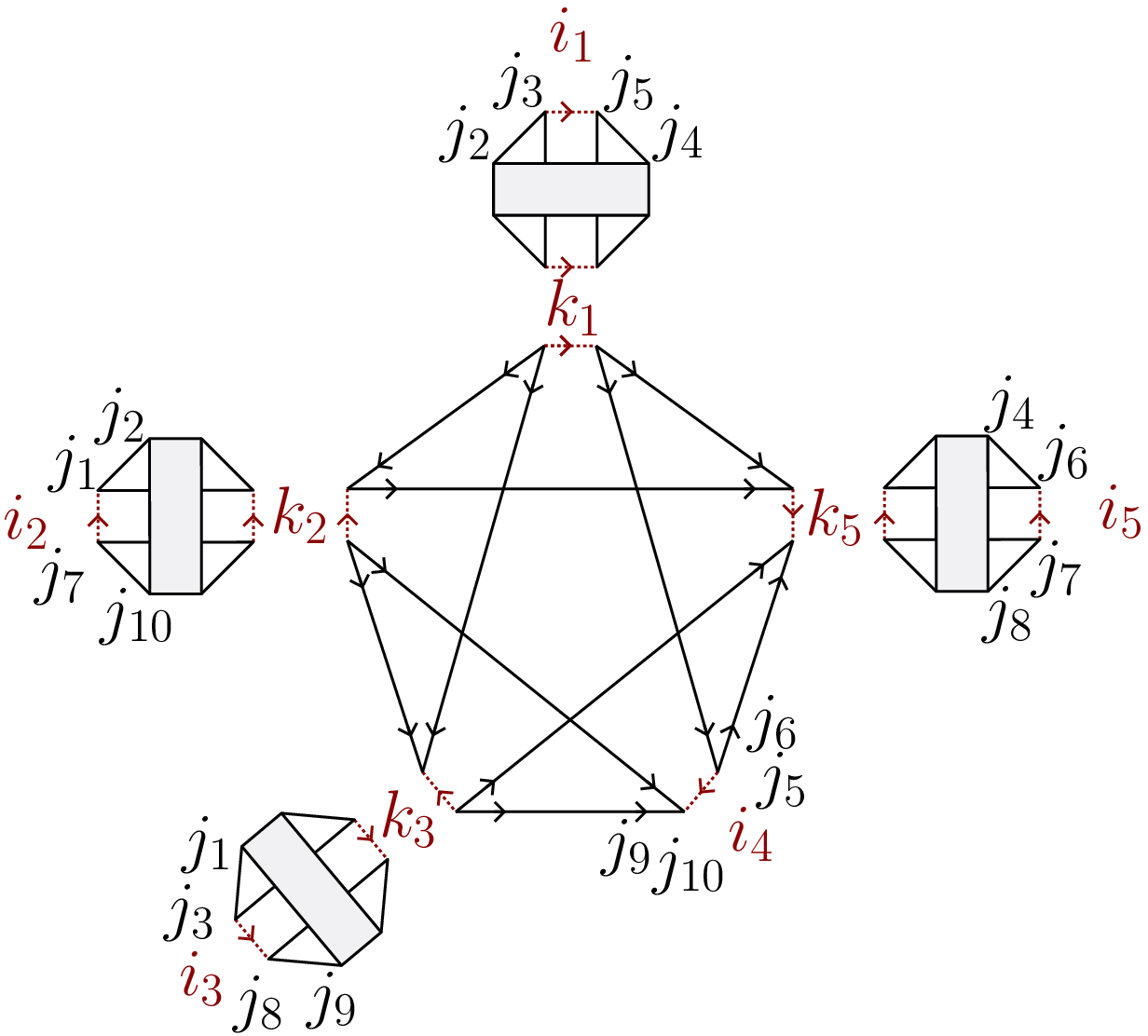}}
 ,
\end{align}
resulting in a superposition of $SU(2)$ $\{15j\}$ symbols weighted by one booster functions $B_{4}$ per edge $e$ in the considered vertex. To obtain this form each $SL(2,\mathbb{C})$ integral is decomposed in a SU(2) ``rotation'', a boost along a fiducial axes and another SU(2) ``rotation''. The compact integrals are then evaluated exactly composing the SU(2) invatiants in \eqref{eq:vertexEPRL}.

The summation is over a set of auxiliary spins $l_{f}$\footnote{that are effectively magnetic indices respect the group $SL(2,\mathbb{C})$ and are originated by the splitting of two representation matrices} for each face involving the vertex but excluding the gauge fixed edge for a total of $6$ distinct $l_f$, with lower bound $l_{f}\geq j_{f}$, and a set of auxiliary intertwiners $k_{e}$ for each edge in the vertex excluding the gauge fixed one for a total of $4$ that can assume all the values compatible with triangular inequalities.

The expression \eqref{eq:vertexEPRL} is far more convenient than the original one for numerical evaluation. We decomposed the amplitude into smaller building blocks, which are easier to compute. We trade four six dimensional unbounded integrals with a summation of SU(2) invariants, which are widely studied, and a class of one-dimensional integrals that we need to perform and we can characterize independently.

These formulas can be generalized to arbirtary valency, for more details see \cite{Sarno:2018ses}. In the library \coolname, in addition to the simplicial amplitude, we also provide the tools to evaluate amplitudes of four valent vertices. The reader can be unfamiliar with the $B_4$ functions, these are the non compact remants of the $SL(2,\mathbb{C})$ integrals and they encode all the details of the EPRL model. We named them booster functions and are defined in the following way: 
\begin{equation}
\label{eq:boosterdef}
B_4(j_f,l_f;i,k) = \frac{1}{4\pi} \sum_{p_a} \left(\begin{array}{c} j_f \\ p_f \end{array}\right)^{(i)} \left(\int_0^\infty \mathrm{d} r \sinh^2r \, 
\prod_{a=1}^n d^{(\gamma j_f,j_f)}_{j_f l_f p_f}(r) \right)
\left(\begin{array}{c} l_f \\ p_f \end{array}\right)^{(k)}\  = \raisebox{-12.5mm}{ \includegraphics[width=2.2cm]{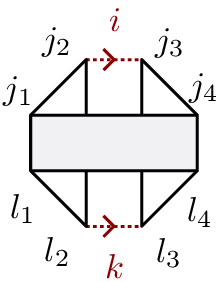}},
\end{equation}
where $d^{(\rho,k)}(r)$ are the boost matrix elements for $\gamma$-simple irreducible representations of $SL(2,\mathbb{C})$ in the principal series, $\gamma$ is the Immirzi parameter and the $(4jm)$ symbols are reported in Appendix \ref{app:su2symbols}. All the test and computations in this paper are performed with a conventional value of $\gamma=1.2$.
The explicit form of the boost matrix elements can be found in the literature in its general form \cite{Ruhl:1970,Naimark:2014,Rashid:1979xv,Speziale:2016axj}, here we just report them in the case of simple irreducible representation:
\begin{align}
\label{eq:dsmall}
d^{(\gamma j,j)}_{jlp}(r) =&  
(-1)^{\frac{j-l}{2}} \frac{\Gamma\left( j + i \gamma j +1\right)}{\left|\Gamma\left(  j + i \gamma j +1\right)\right|} \frac{\Gamma\left( l - i \gamma j +1\right)}{\left|\Gamma\left(  l - i \gamma j +1\right)\right|} \frac{\sqrt{2j+1}\sqrt{2l+1}}{(j+l+1)!}  
\left[(2j)!(l+j)!(l-j)!\frac{(l+p)!(l-p)!}{(j+p)!(j-p)!}\right]^{1/2} \nonumber \\
&\ \times e^{-(j-i\gamma j +p+1)r}
\sum_{s} \frac{(-1)^{s} \, e^{- 2 s r} }{s!(l-j-s)!} \, {}_2F_1[l+1-i\gamma j,j+p+1+s,j+l+2,1-e^{-2r}] \ .
\end{align}
The study of the booster function is still in the initial stage, more work is necessary to fully understand all their properties. For example, while the symmetries are known, the recursion relations still need to be explored. The study of their semiclassical limit is in a preliminary stage \cite{citaPierre} but an appealing geometrical picture seems to emerge.  

On one hand, the introduction of booster functions brings a drastic simplification to the computation of the vertex amplitude because it reduces the problem of dealing with many high oscillatory integrals to the study a family of one dimensional integrals, which are easier to handle and manipulate. On the other hand, we introduce a large amount of convergent summation in the computation which need to be dealt with. The net gain of this operation is evident once we realize that each element of the sum can be computed in a fast way and with high precision.

\section{The \coolname library}
\label{sec:library}
\subsection{General strategy}
\label{sec:strategy}
We can summarize the evaluation of the vertex amplitude \eqref{eq:vertexEPRL} in three main ingredients that we need to mix in the right way:
\begin{itemize}
\item all the necessary $\{15j\}$ symbols,
\item all the required booster functions,
\item the sum over the auxiliary spins and intertwiners.
\end{itemize}

In the next sections we will discuss in great detail the problems  we encounter in these three steps and their possible solutions. Here we want to give a general overview. The first valuable lesson we learned is to not waste time computing vanishing contributions. Many SU(2) invariants and booster functions are zero because of symmetries (triangular inequalities, odd symmetries under permutations), we avoid computing objects that are part of amplitudes that we can identify as vanishing a priori. 

Symmetries also play a secondary but equally important role. The objects in the summation repeat themselves, instead of wasting time doing the same computation over and over again, we store them in the RAM and, at the end, we dump them to disk. To make it possible and not saturate the memory in a couple of iterations, we optimize the computation factoring the symmetries and consider only one representative per equivalence class. The bottleneck of the computation while dealing with SU(2) invariants is not the evaluation time but the memory needed. In an equivalence class of $\left\lbrace 6j\right\rbrace$ symbols there are approximately a hundred elements, we store only one representative per class reducing by two orders of magnitude the amount of necessary memory.

When dealing with summations with a huge amount of terms we need to be extremely careful. In our calculation, apart from the summation over auxiliary spins and intertwiners itself, these sums appear also in the evaluation of the booster functions. In particular, to deal effectively with interference effects we need to resort to arbitrary precision libraries \cite{GMP,MPFR,MPC}. 
How to sum a large number of finite precision floating point numbers is a well-studied problem in numerical analysis. The bound of the worst-case error on the sum of $N$ numbers with floating point precision grows proportional to $N$. By using a compensated summation algorithm \cite{Kahan} we can make this error independent of $N$, and having it depend only on the floating-point precision of the addends.
Moreover, we employ compensated summation algorithms to significantly reduce the numerical error in the addition of finite precision floating point numbers. 

The full computation is divided into a large number of small tasks and we can efficiently parallelize it with \texttt{OpenMP}. We parallelize the computation of the booster functions and the summation over the auxiliary spins but not the evaluation of the SU(2) invariants. The computational time needed for the evaluation of the SU(2) invariants is negligible respect to all the other steps.

The sums over the six auxiliary spins are unbounded and convergent, therefore to extract a number from them we need to truncate. We introduce a homogeneous cutoff on all the spins $j_f \leq l_f \leq j_f + \Delta$ and we sum over all compatible intertwiners $k_e$.
The details of the convergence of this summation strongly depend on the value of the face spins and edge intertwiners, hindering our ability to provide a consistent estimate of the systematical error due to the truncation. We are then forced to perform an analysis a posteriori: we report some examples in Figure \ref{fig:4SimplexConvergence} where it is possible to estimate it between $1$ and $5$ percent for $\Delta\approx5$.

We mention here that we studied the possibility to develop an algorithm to guess an appropriate value for the cutoff as a function of the desired precision. However, we decided not to implement it in the current version of the library because we are lawing a way to deal with non monotonic convergences. 

The error on the booster functions, due to the discretization of the integral, and the one on the SU(2) invariants are completely irrelevant with respect to the one due to the truncation. We will discuss them in more detail in the following sections.

For a single 4-simplex with fixed intertwiners, the bottleneck of the computation is the evaluation of the booster functions. If we need to compute the amplitude for all the possible intertwiners, for example, because we are contracting more than a single vertex, most of the computational time is used in performing the summation. Having in mind the goal of performing the computation of spin foam diagrams with many vertices, since the number of elements grows exponentially with it, we are studying selection rules based on the asymptotic of the booster functions or with the machine learning techniques. 

%----------------------------------------------------------------------------
\subsection{SU(2) invariants}
\label{sec:su2}
Evaluating a significant amount of SU(2) invariants can be a resources draining task. An SU(2) invariant is defined as a summation over the magnetic indices of intertwiners (and them as a sum of $(3jm)$ symbols). Their evaluation, directly from the definition, is not convenient, as it will exhaust too many computational resources, both in time and memory. To compute an amplitude we need to evaluate a large number of invariant, requiring a technique that is both fast and memory lightweight. We can solve the time issue by reducing to one the number of summation. We can reduce every SU(2) invariant in a sum of $\{6j\}$ and $\{9j\}$ symbols that can be directly computed using prebuilt libraries \cite{Johansson:2015cca}. We tackled this problem for the first time in \cite{Dona:2017dvf} where the asymptotic of a general SU(2) invariant has been studied analytically and also numerically in the simplicial case. We employ a similar technique in \coolname. Following the notation of the monography \cite{Yutsis:1962vcy} we identify five different kinds of $\{15j\}$ symbols. 
For example, we can write the  $\{15j\}$ symbols of the first kind as a sum of the product of five $\{6j\}$s symbols:
\begin{align}\label{eq:15jIrr}
\{15j\}_I := \raisebox{-15mm}{\includegraphics[width=3.5cm]{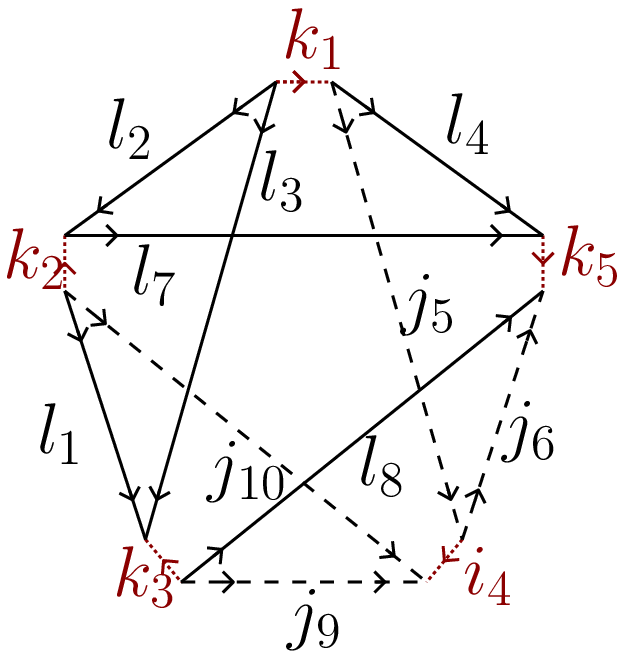}} &= \sum_x (2 x +1) (-1)^{\sum_i l_i + j_i + k_i + i_4 }
\Wsix{k_1}{l_7}{x}{k_5}{j_5}{l_4} \Wsix{j_5}{k_5}{x}{l_8}{i_4}{l_6} \\ \nn & \times \Wsix{i_4}{l_8}{x}{k_3}{j_{10}}{j_9} \Wsix{j_{10}}{k_3}{x}{l_3}{k_2}{l_1} \Wsix{k_2}{l_3}{x}{k_1}{l_7}{l_2}.
\end{align}
To compute a single amplitude at a fixed cutoff $\Delta$, we need to evaluate an enormous amount of $\{15j\}$ symbols, and consequently of $\{6j\}$ (or  $\{9j\}$) symbols. 
We recall that all the spins $l_f$ vary in the range between the face spin  $j_f$ and the cutoff $j_f+\Delta$, while the intertwiners $k_e$ assume all the values allowed by the triangular inequalities. %Even if this looks like a simple task, we encounter two principal difficulties.

The development of algorithms that allow for a fast and accurate computation of $\{6j\}$ symbol is a fascinating and challenging topic studied by mathematicians and computer scientist, and a significant amount of libraries for the job are available. In this work, we adapt the C library \texttt{wigxjpf} developed in \cite{Johansson:2015cca} that can be used to compute  $(3jm)$, $\{6j\}$ and $\{9j\}$ symbols. We refer to their original paper for a detailed discussion of the library's performances, accuracy and memory management.

We also solved the memory usage issue by efficiently storing $\{6j\}$ symbols taking into account the various symmetries of the symbols. The right data structure for this job is the hash table, a structure that efficiently maps keys to values. We associate to each $\{6j\}$ symbol a key that is automatically shared by the whole class of symbols with the same value (because of symmetries). Every $\{6j\}$ symbols share its value with other $143$ \cite{Rasch:2003} symbols due to permutation symmetries.
Notice that an extension of the library \texttt{wigxjpf} named \texttt{fastwigxj} provides a hash table implementation for $(3jm)$, $\{6j\}$ and $\{9j\}$ symbols but, since it is designed for more general applications, it stores all possible combinations of the desired invariants with \textit{all} spins smaller than a cutoff. This amount of data is way more of what we need in an ordinary EPRL four simplex computation resulting in a huge memory consumption, and consequently being an obstacle when exploring larger values of the spins. To circumvent this limitation, we implement our own hash tables based on \texttt{khash} provided in the library  \texttt{Klib} \cite{Klib2016}. We generate the keys of the hash table with the hash functions discussed in \cite{Rasch:2003}, and we store the $\{6j\}$ symbols assigning a unique six integers key to each symmetry class, saving one representative value. 

We can introduce an additional simplification to improve the performance, as already used in \cite{Dona:2017dvf}. We can use the freedom of choosing the intertwiners recoupling basis of the spin foam edges to rewrite the $\{15j\}$ symbols as:
\begin{align} \label{eq:15JRed}
\{15j\}_R := \raisebox{-15mm}{\includegraphics[width=3.5cm]{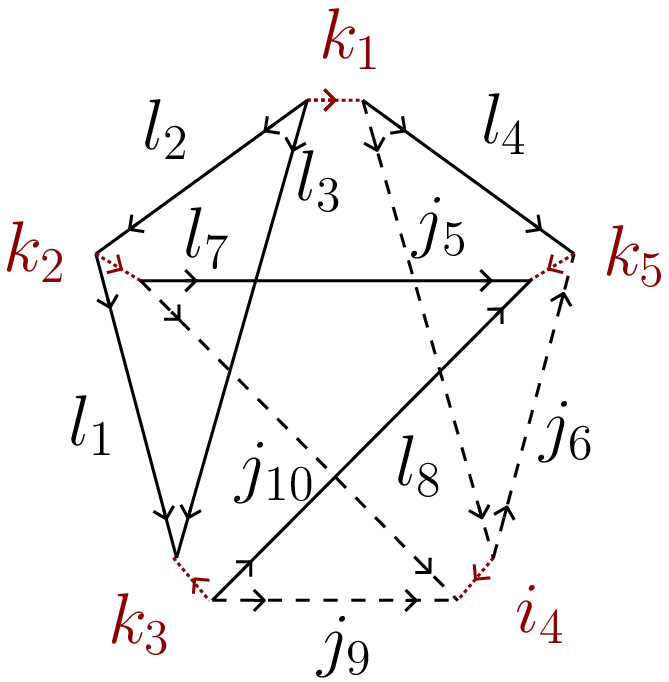}} &= (-1)^{2 (l_4+l_3) + 2k_1 - (k_2 + k_3 + i_4 + k_5) } 
 \\ \nn & \times \Wsix{k_1}{k_3}{k_2}{l_1}{l_2}{l_3} \Wsix{k_1}{i_4}{k_5}{j_6}{l_4}{j_5}  \Wnine{k_2}{k_3}{k_1}{j_{10}}{j_9}{i_4}{l_7}{l_8}{k_4} \ ,
 \end{align}
and then rewrite the  $\{9j\}$ symbol as a sum of three $\{6j\}$ symbols:
\begin{align} \label{eq:15JSplit}
\{15j\}_R &=  (-1)^{2 (l_4+l_3) + 2k_1 - (k_2 + k_3 + i_4 + k_5) } \Wsix{k_1}{k_3}{k_2}{l_1}{l_2}{l_3} \Wsix{k_1}{i_4}{k_5}{j_6}{l_4}{j_5}  \\ \nn & \times\sum_x (2x+1) (-1)^{2x} \Wsix{k_2}{j_{10}}{l_7}{l_8}{k_5}{x} \Wsix{k_3}{j_9}{l_8}{j_{10}}{x}{i_4} \Wsix{k_1}{i_4}{k_5}{x}{k_2}{k_3} \ .
\end{align}
Comparing with \eqref{eq:15jIrr} we are still using five $\{6j\}$ symbols but we are summing only over three of them, resulting in a relevant save of time and memory. 
This simplification is quite in general, (is free if the edges are not boundary ones), but we need to be careful in computations of spin foams transition amplitudes with many vertices, since the base choice on one edge affects both vertices it connects, and the split \eqref{eq:15JSplit} needs to be done consistently in every vertex. For an explicit example of a five vertex transition amplitude where this splitting is performed see Section 4.4 of \cite{Dona:2018pxq}.

\begin{figure}[H]
\begin{center}
\includegraphics[width=\textwidth]{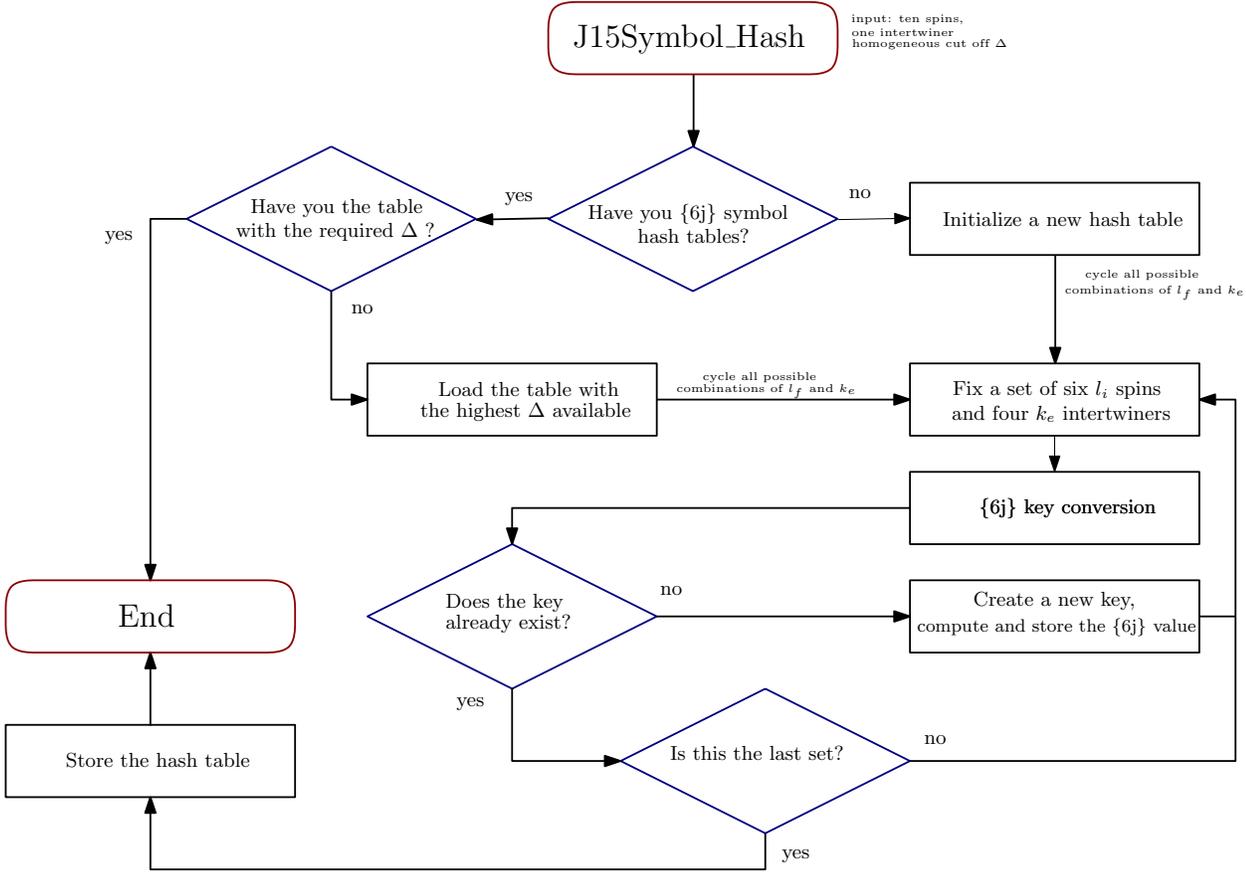}
\end{center}
\caption{\texttt{J15Symbol\textunderscore Hash} is the main function to store in an efficient way $\{6j\}$ symbols.}
\label{fig:FlowChart_J15Hash} 
\end{figure}

\begin{figure}[H]
    \centering
    \begin{subfigure}[b]{0.49\textwidth}
        \includegraphics[width=7.5cm]{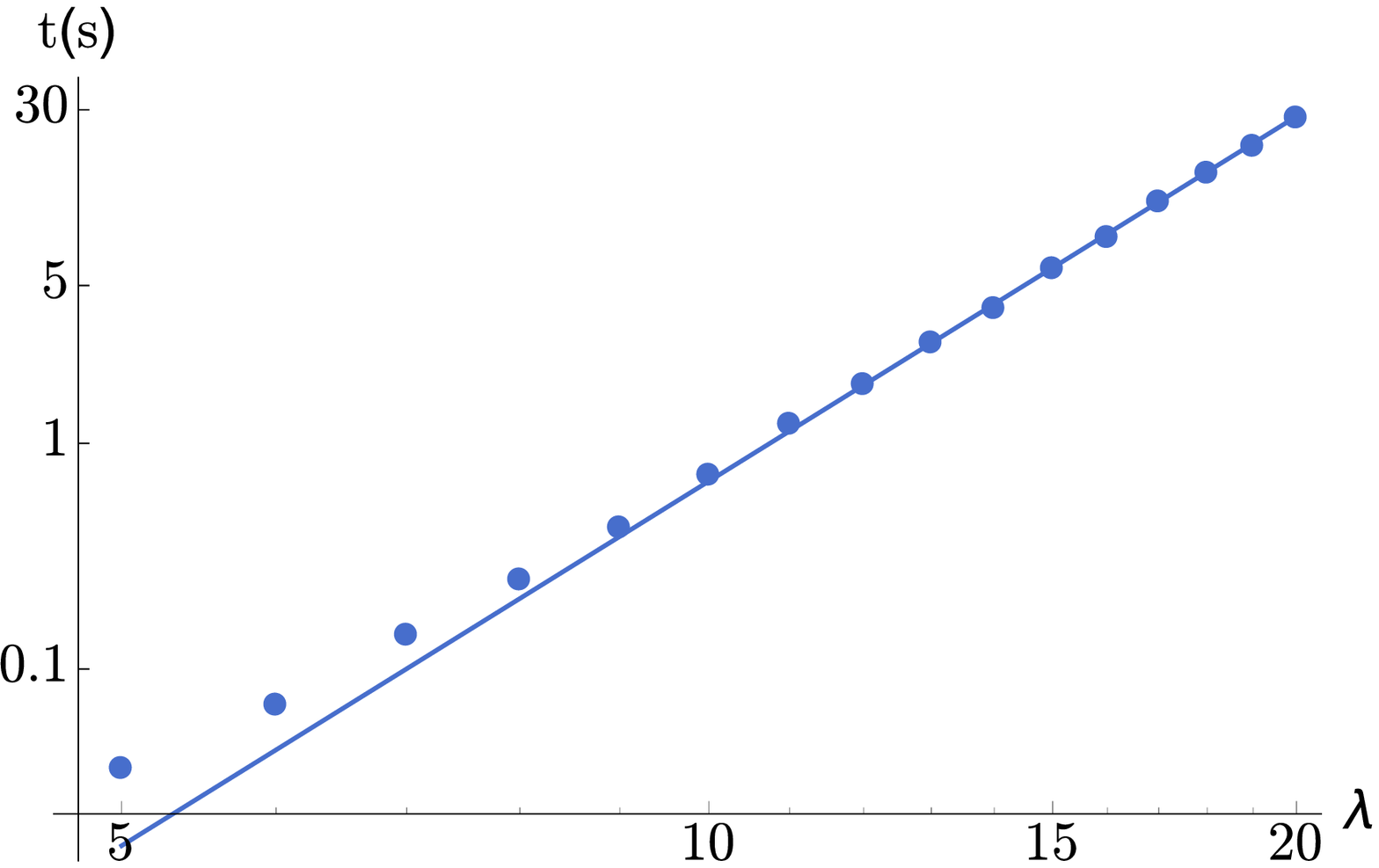}
    \end{subfigure}
    \begin{subfigure}[b]{0.49\textwidth}
        \includegraphics[width=7.5cm]{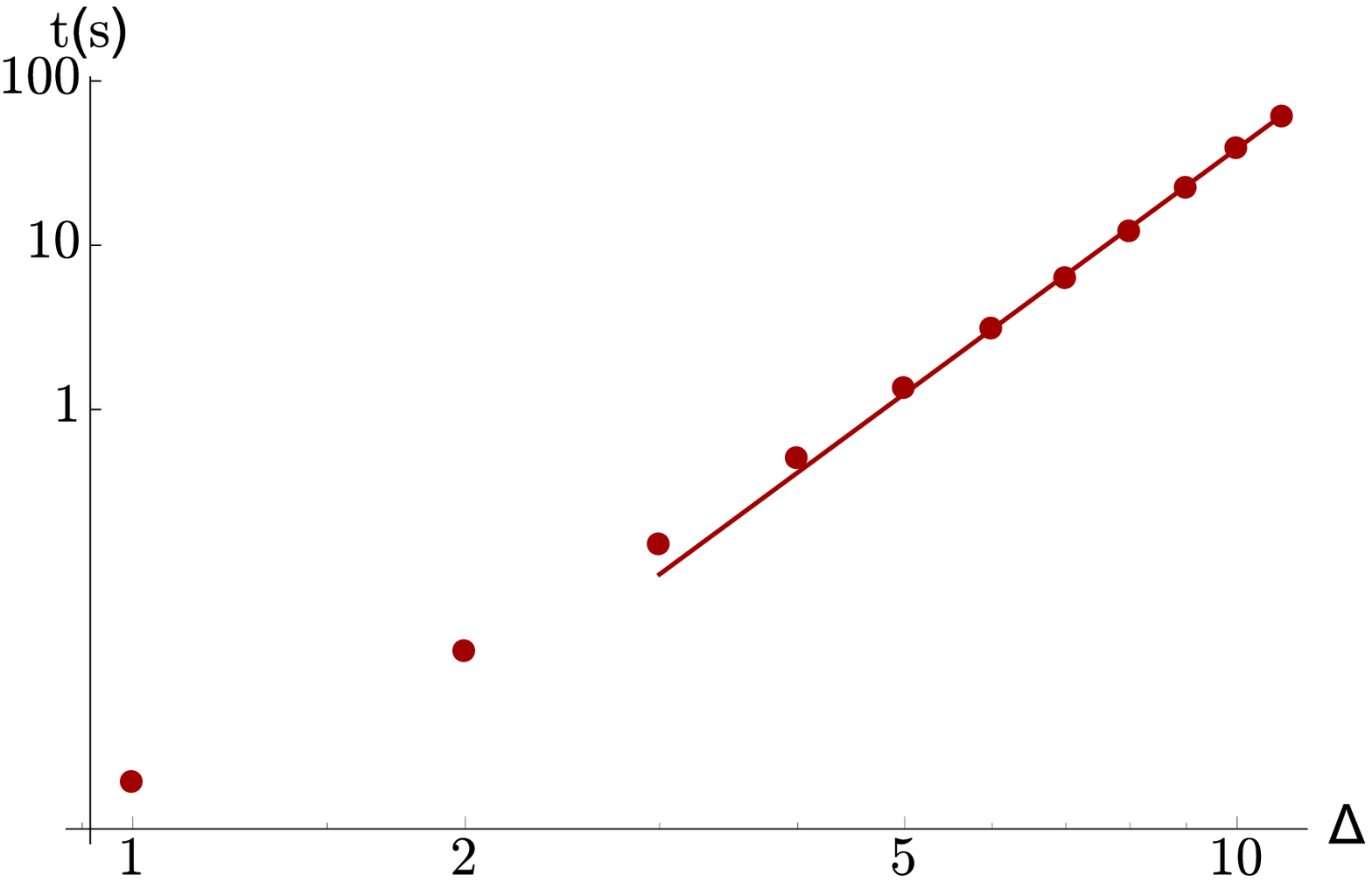}
    \end{subfigure}
\caption{\label{fig:15JTiming} 
\small{\emph{ Evaluation time of the function} \texttt{J15Symbol\textunderscore Hash}. Left panel: \emph{Scaling of the configuration with all the spins and the gauge fixed intertwiner equal $j_f=i_4=\lambda$ and a vanishing cutoff $\Delta=0$, we compute the $\{15j\}$ symbols for all the values of the intertwiners $k_e$. We see a power law trend $a\times j^b$ with $a=2.9\times 10^{-6} s$ and $b=5.3$. } Right panel: \emph{Scaling in the cutoff of the configuration with all the spins and the gauge fixed intertwiner equal to $j_f=i_4=1$. We see a power law trend $a\times j^b$ with $a= 4\times 10^{-4}s$ and $b= 5.0$.} } }
\end{figure}

\begin{figure}[H]
    \centering
    \begin{subfigure}[b]{0.49\textwidth}
        \includegraphics[width=7.5cm]{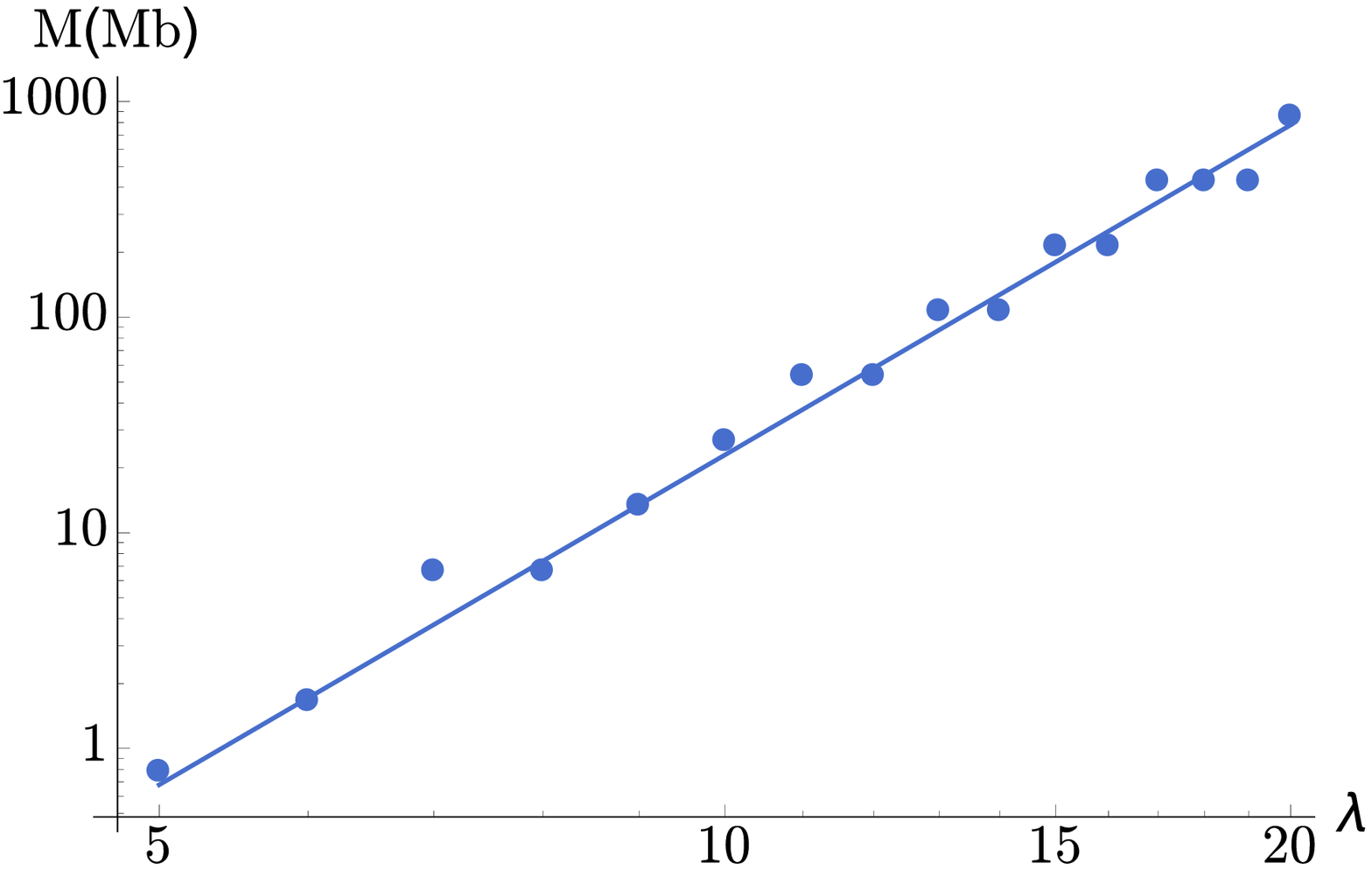}
    \end{subfigure}
    \begin{subfigure}[b]{0.49\textwidth}
        \includegraphics[width=7.5cm]{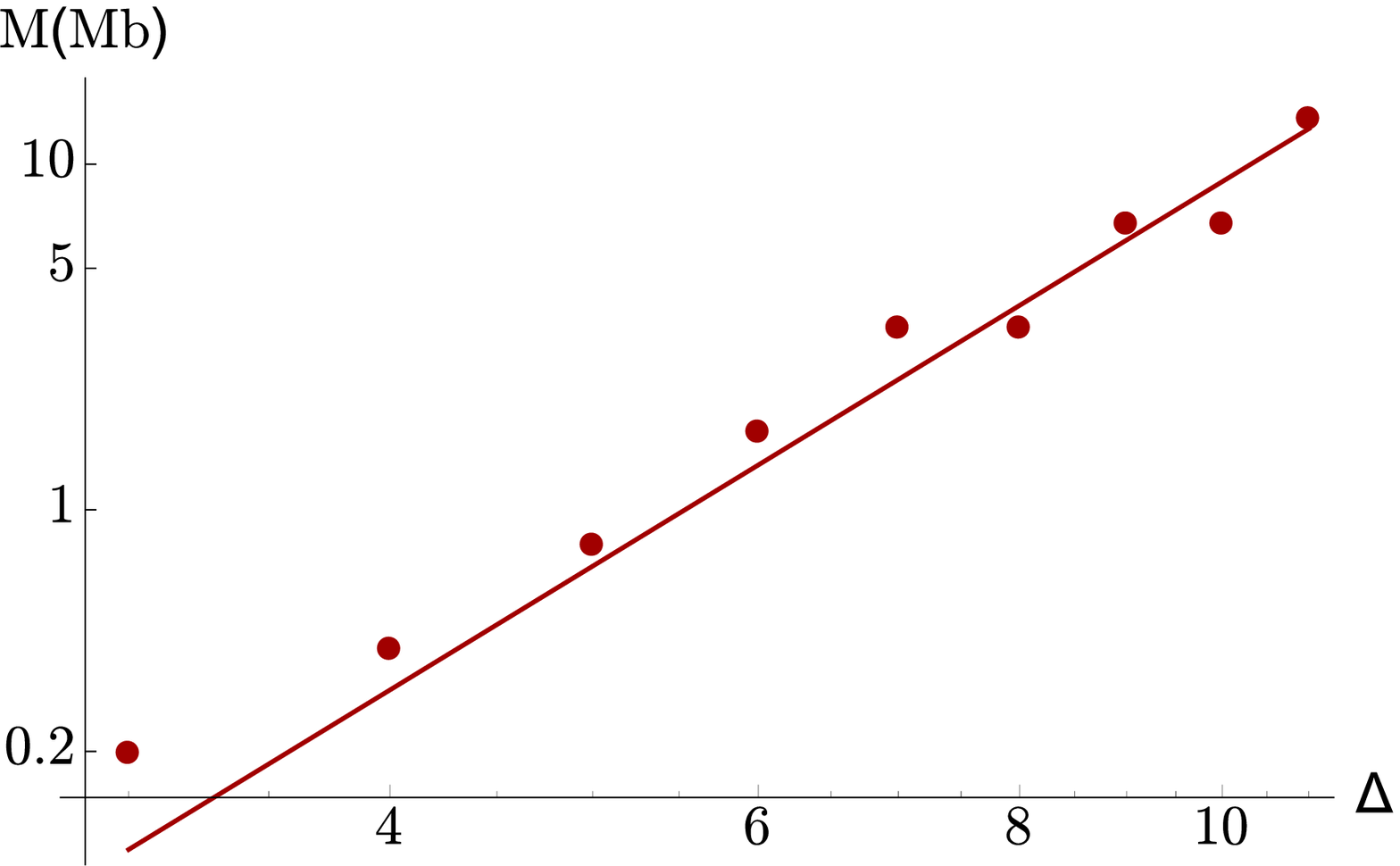}
    \end{subfigure}
      \caption{\label{fig:15JMemory} 
      \small{\emph{
      Memory usage of the function} \texttt{J15Symbol\textunderscore Hash}. Left panel: \emph{Scaling of the configuration with all the spins and the gauge fixed intertwiner equal $j_f=i_4=\lambda$ and a vanishing cutoff $\Delta=0$, we compute the $\{15j\}$ symbols for all the values of the intertwiners $k_e$. We see a power law trend $a\times j^b$ with $a=1.8\times 10^{-4} \mathtt{Mb}$ and $b=5.1 $. } Right panel: \emph{Scaling in the cutoff of the configuration with all the spins and the gauge fixed intertwiner equal to $j_f=i_4=1$. We see a power law trend $a\times j^b$ with $a=1.8\times 10^{-3}  \mathtt{Mb}$ and $b= 3.7$.} } }
\end{figure}

The function where we implemented the evaluation and the storing of the $SU(2)$ invariants is called \texttt{J15Symbol\textunderscore Hash}. We sketch its working principles in the flowchart in Figure \ref{fig:FlowChart_J15Hash}. 
The function prepares all the $\{6j\}$ symbols in \eqref{eq:15JSplit} needed for the computation of a vertex amplitude given the 10 spins $j_f$ $f=1,\ldots,10$ and a cutoff $\Delta$. Following the conventions of \eqref{eq:15JSplit} the gauge fixed edge is identified by the four spins $j_5$, $j_6$, $j_9$ and $j_{10}$ in the recoupling basis $\left(j_5,j_6\right)$ with intertwiner $i_4$. After the evaluation, the hash tables are stored in a file indexed with the name of the gauge fixed edge and the cutoff $\Delta$, with the naming convention \texttt{2j5.2j6.2j9.2j10\textunderscore 2i4\textunderscore 2$\Delta$}. The factor of two multiplying the spins deals with the possible half-integer values. We decided not to use all boundary variables in the indexing to avoid files proliferation when computing amplitudes with unbounded internal faces. With our convention, all the needed symbols will be stored in the same file. In this way, we can reach a compromise between hash tables' memory consumption and time necessary to read and write the hash tables on disk. 
We strongly suggest selecting the gauge fixed edge as part of a boundary face to minimize the number of files created. We report a typical time and memory consumption of \texttt{J15Symbol\textunderscore Hash} in Figure \ref{fig:15JTiming} and \ref{fig:15JMemory}.


As a possible upgrade for the future, we are planning to avoid the splitting of the $\{9j\}$ symbol as a sum of three $\{6j\}$ symbols in \eqref{eq:15JSplit} and evaluate it directly with \texttt{wigxjpf} with consequent hashing and storing. This upgrade hasn't been implemented yet since it requires an efficient way to assign a unique key to each symmetry class of the $\{9j\}$ symbol, that we are still studying. Anyway, while this upgrade will significantly improve the speed of low spin computations, it will not be applicable in regimes with larger spins because it would use too much memory.

  %----------------------------------------------------------------------------
\subsection{Booster functions}
\label{sec:booster}
The expression for the booster functions \eqref{eq:boosterdef} requires the calculation of the integral over the rapidity of four boost matrix elements \eqref{eq:dsmall} and a sum over magnetic indices. The latter is not a problem in this case since the number of terms to be summed over is relatively small (for all equal spins $j$ is less than $(2j)^3$ terms). However, the numerical integration is still very challenging. The integrand is a superposition of highly oscillating hypergeometric functions that, when multiplied together, interfere resulting in a localized, smooth and regular function. Moreover, for specific values, the hypergeometric functions are evaluated near one of their poles. These apparent divergences will cancel with other terms in the superposition but would require extremely high precision to control the accuracy of the result. We solved these issues by attacking them from a different angle.
A direct computation is very slow, and with this motivation an alternative formula was proposed in \cite{Speziale:2016axj} and numerically explored by \cite{citaGozzini}. Nevertheless, the situation improves significantly using a different representation for the boost matrix elements \eqref{eq:dsmall}. This problem has been explored in \cite{citaFrancois} where each boost matrix elements \eqref{eq:dsmall} is rewritten as a finite sum of exponentials with complex coefficients. For simple irreducible representations and $\gamma \neq 0$, the result is given by:  
\begin{align}
d^{(\gamma j,j)}_{jlp}(r) &= 
 \frac{1}{(e^{r}-e^{-r})^{j+l+1}} \\ &  \nn
\times  \left [ \sum_{m=0}^{j+l - | j-p|} Y_m^{\gamma j, j; j,l;p}e^{(j+l- | j-p | -2m -i\gamma j)r}+(-1)^{l-j} \sum_{n=0}^{j+l - | j-p|} \overline{Y}_n^{\gamma j, j; j,l;-p}e^{(j+l- | j-p | -2n +i\gamma j)r} \right ],
\end{align}
where $Y$ are complex coefficients \footnote{We refer to \cite{citaFrancois} for their full expression.}. In the same work they provide a C++ implementation of the booster functions based on this formula. We integrated this code in \coolname porting it in C, we added the evaluation of the $(4jm)$ symbols in \eqref{eq:boosterdef} with \texttt{wigxjpf} and we provided an hash table data structure similar to the one we introduced for the SU(2) invariants in the previous section. The evaluation of these functions requires arbitrary precision floating point numbers, implemented with \texttt{MPFR} and \texttt{MPC}. These tools allow us to compute the booster functions in an accurate way and to test some of theirs properties.

\begin{figure}[H]
    \centering
    \includegraphics[width=7.5cm]{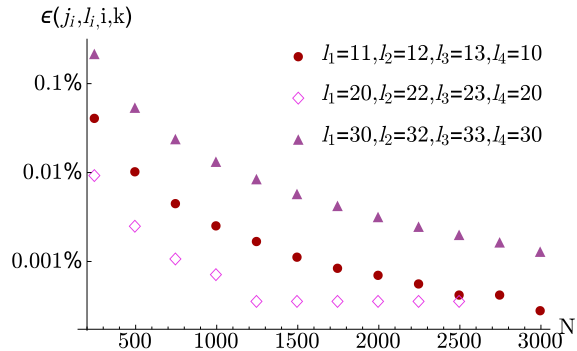}
    \caption{\label{fig:B4Error} \small{\emph{Dependence of the booster function on the number of points in the rapidity integral. We plot the relative difference $\epsilon$ with respect to the value with $10^4$ points. All three booster functions reported here have all equal spins $j_f=10$ and intertwiner $i=0$. The four spins $l_f$ and intertwiner $k$ are respectively: $l_f=11,12,13,10$ and $k=1$, $l_f=20,22,23,20$ and $k=2$, $l_f=30,32,33,30$ and $k=2$.} } }
\end{figure}

The integral is performed using the trapezoidal rule with 3000 points as a default number. We explored this choice in Figure \ref{fig:B4Error}. The convergence looks very fast in the number of points; we observe a slight worsening with the increase of the $l_f$. The number of integration points can be increased arbitrarily to reduce even further the admissible error on the single booster function.

\begin{figure}[!ht]
\begin{center}
\includegraphics[width=\textwidth]{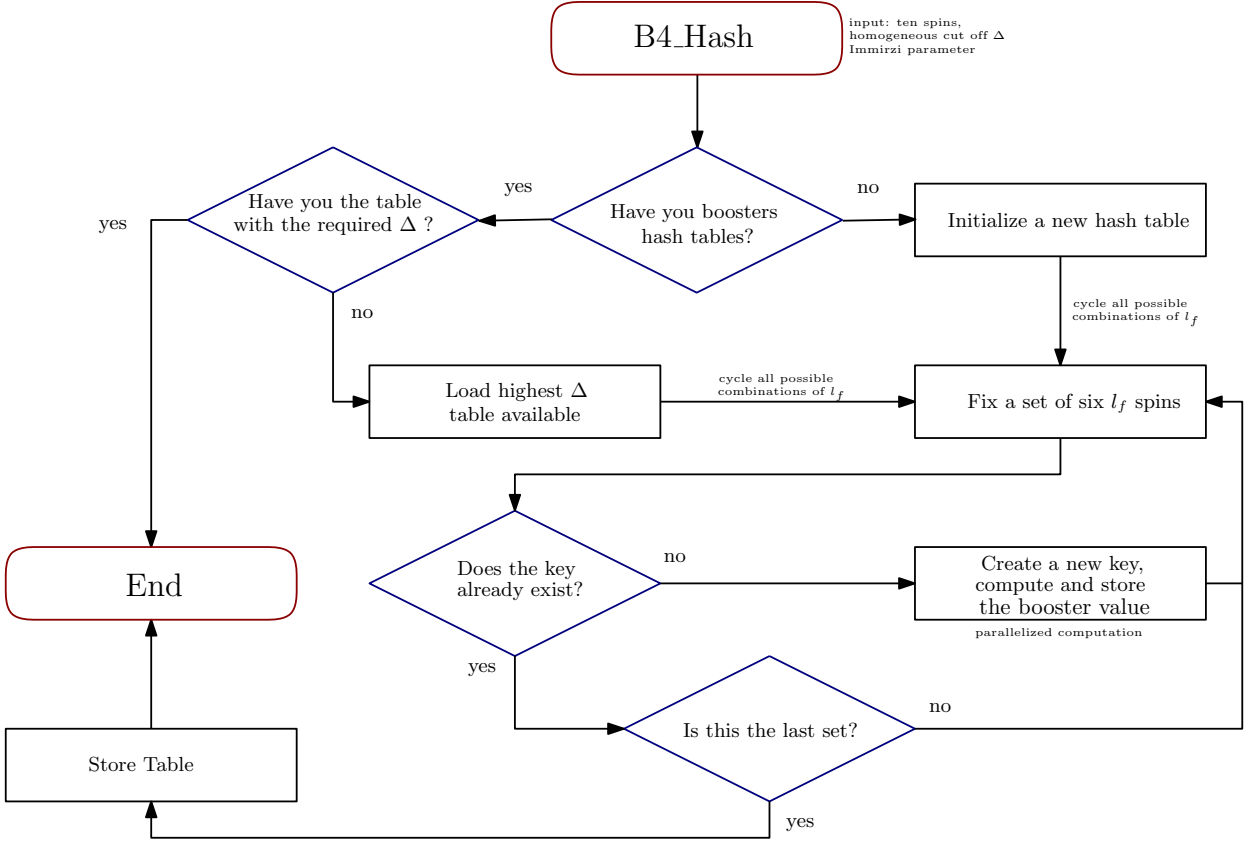}
\end{center}
\caption{ \texttt{B4\textunderscore Hash} compute and store efficiently the booster functions.}
\label{fig:FlowChart_B4Hash} 
\end{figure}

The sampling of the integration interval is parallelized with \texttt{OpenMP}. The library \coolname implement the evaluation of all the booster functions needed in the computation of a vertex amplitude with the function \texttt{B4\textunderscore Hash}. We illustrate its structure in the diagram in Figure \ref{fig:FlowChart_B4Hash}.

\begin{figure}[H]
    \centering
    \begin{subfigure}[b]{0.49\textwidth}
        \includegraphics[width=7.5cm]{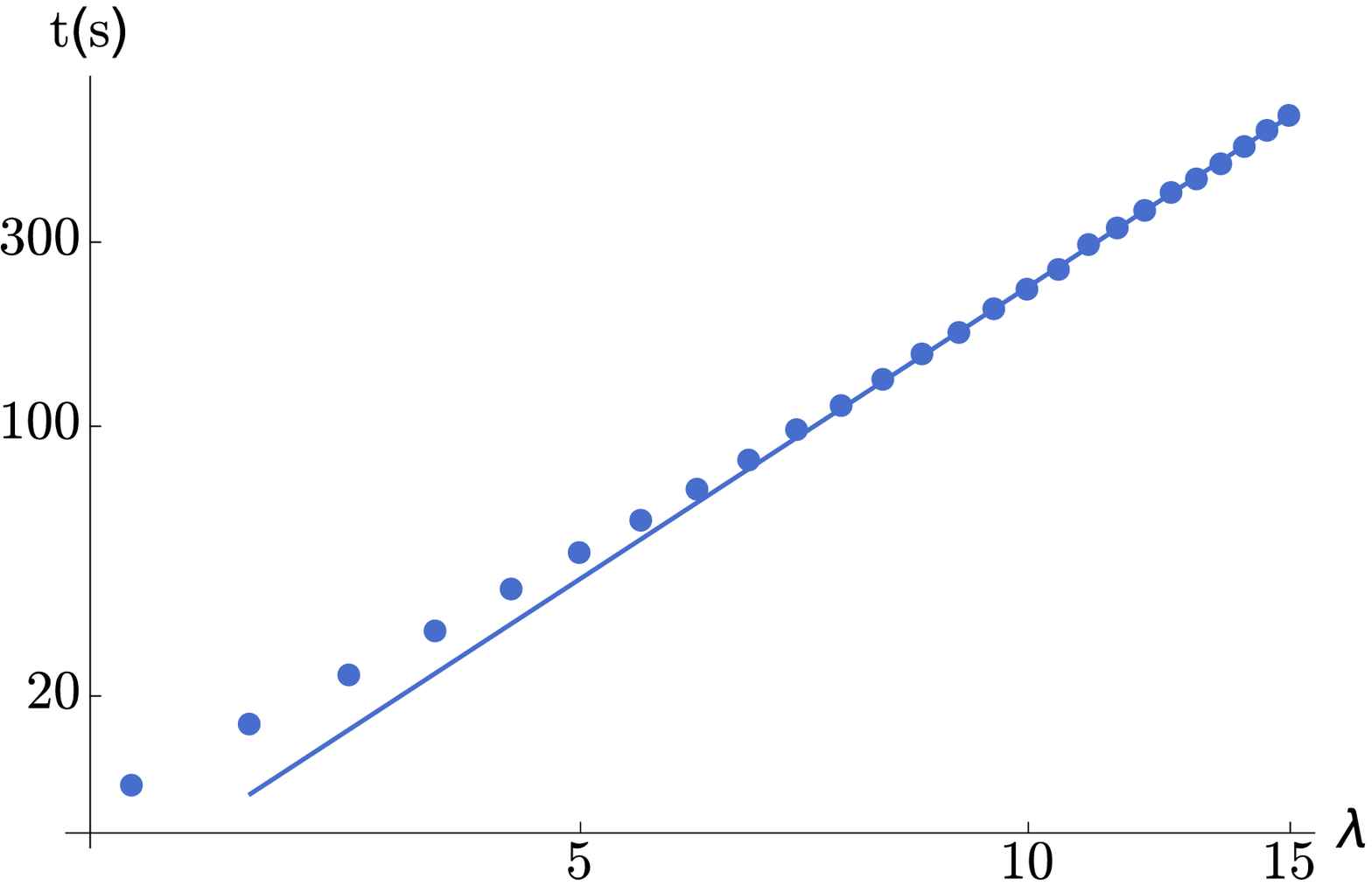}
    \end{subfigure}
    \begin{subfigure}[b]{0.49\textwidth}
        \includegraphics[width=7.5cm]{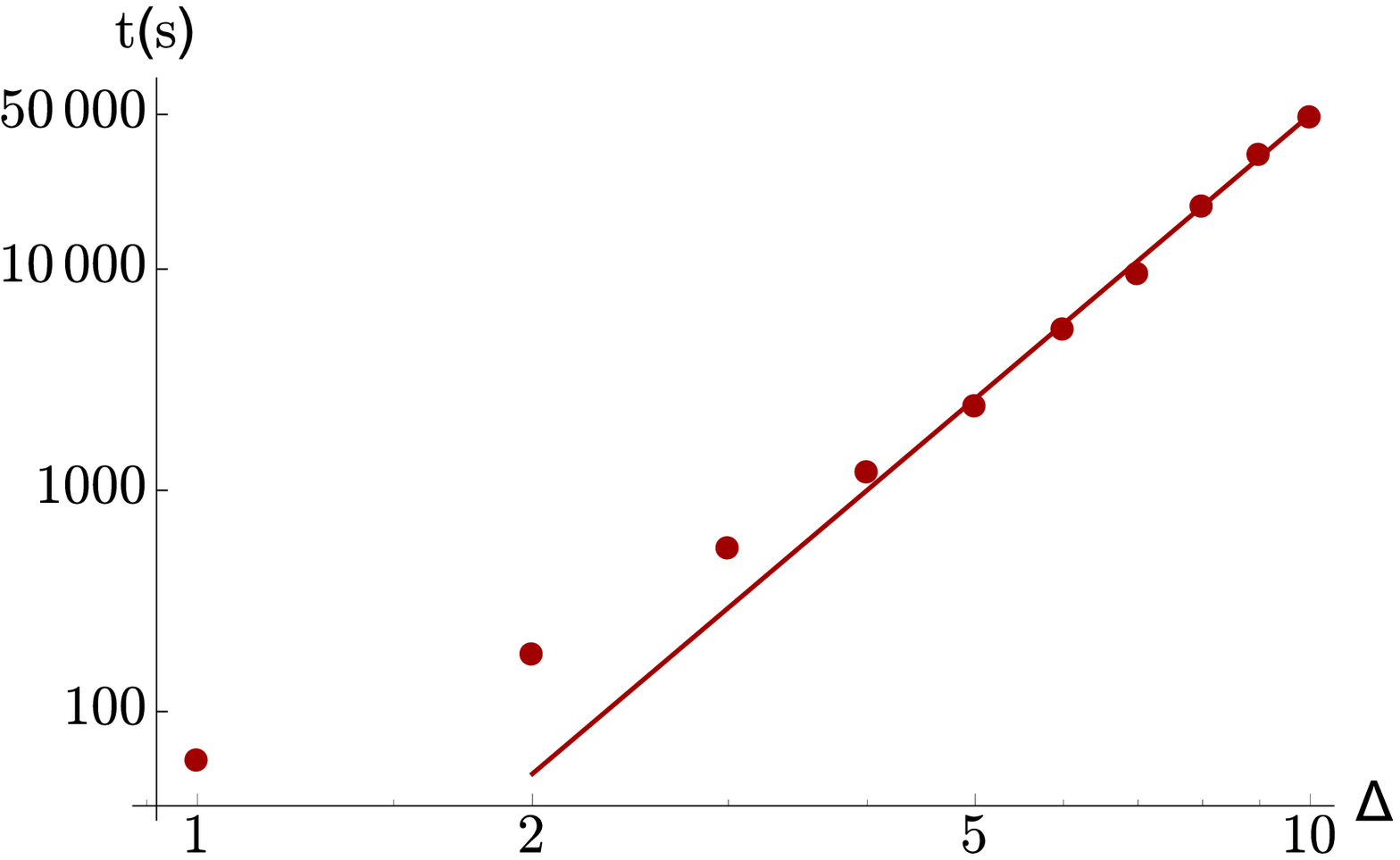}
    \end{subfigure}
      \caption{\label{fig:B4timing} \small{\emph{  
Evaluation time of the function} \texttt{B4\textunderscore Hash}. Left panel: \emph{Scaling of the configuration with all the spins and the gauge fixed intertwiner equal $j_f=i_4=\lambda$ and a vanishing cutoff $\Delta=0$. We compute all the booster functions for all the values of the intertwiners $k_e$. We see a power law trend $a\times j^b$ with $a=0.7 s$ and $b=2.5$. } Right panel: \emph{Scaling in the cutoff of the configuration with all the spins and the gauge fixed intertwiner equal to $j_f=i_4=1$. We see a power law trend $a\times j^b$ with $a=2.7s$ and $b= 4.3$.} } }      
\end{figure}

\begin{figure}[H]
    \centering
    \begin{subfigure}[b]{0.49\textwidth}
        \includegraphics[width=7.5cm]{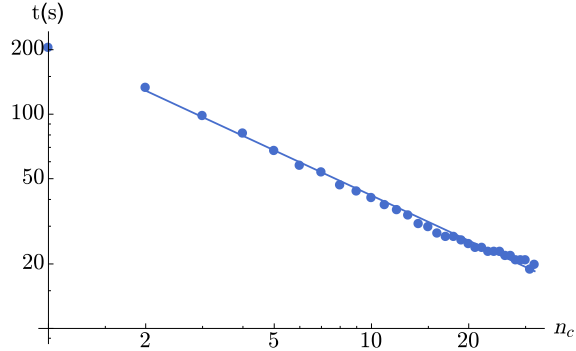}
    \end{subfigure}
      \caption{\label{fig:B4cores} \small{\emph{Parallelization at fixed workload: scaling of the time required by the computation of booster functions for a 4-simplex with $j_f=10$ and $\Delta = 0$ while increasing the number of cores. The power law $a\times j^b$ reads $a=209s$ and $b=-0.7$. } } }\end{figure}

The values of the booster functions are stored in hash tables and then in a file indexed with the name of the gauge fixed edge and the cutoff $\Delta$, using the same naming conventions as the $\{6j\}$ symbols \texttt{2j5.2j6.2j9.2j10\textunderscore 2$\Delta$}.
Furthermore, we divide the files into subfolders depending on the Immirzi parameter $\gamma$ to avoid mixing of incompatible evaluations in case one is interested in working with different values for $\gamma$. We store in each file the booster functions with all the possible combinations of the six $l_f$ spins appearing in a vertex amplitude and every compatible intertwiner. We report a typical time consumption of \texttt{B4\textunderscore Hash} in Figure \ref{fig:B4timing} and the scaling of the time as a function of the number of cores used in the computation in Figure \ref{fig:B4cores}. The memory's use of booster functions is negligible compared to $\{6j\}$ symbols' one.

For completeness, we provide also a function, called \texttt{B4Function}, that computes a single booster, in case one is interested in them outside the context of vertex amplitudes.

When computing the booster functions, one notices that, for a fixed set of spins $j_f$ and $l_f$, not all combinations of intertwiners contributes with values of similar order of magnitude. An example of this can be seen in \cite{Speziale:2016axj} and in \cite{Puchta:2013lza}, for the case $l_f=j_f$ (called simplified model), where the main contribution is given by the same intertwiners $i_e = k_e$. The generalization of this result for any set of $l_f$ is unfortunately still unknown. 
This result, or an equivalent method to localize the dominant contribution from the booster functions, will immediately translate into a significant save of time in the evaluation of the amplitude, especially in the case of multiple vertices.  

We also plan to explore the possibility of completely changing the way in which we compute the booster functions and use their expression in terms of the Clebsch-Gordan coefficients for $SL(2,\mathbb{C})$ introduced in \cite{Speziale:2016axj}.
Using the Clebsch-Gordan formulation of the booster functions, we trade the integrals over the rapidity and the sum over the magnetic indices, with a single integral (and a finite sum) over a unitary $SL(2, \mathbb{C})$ virtual irrep. To do so, one has to compute Clebsch-Gordan coefficients fast and with high accuracy. The speed of the functions we included in \coolname that do this task is too low to compete with the actual implementation of the booster functions. For this reason, we have not implemented it yet. The possibility of using recursion relations fro the Clebsch-Gordan coefficients also makes this option very appealing. A first attempt was made in \cite{citaGozzini}, we think this is a very promising route to improve \coolname.

%----------------------------------------------------------------------------
\subsection{Four simplex amplitude}
\label{sec:amplitude}
\begin{figure}[H]
    \centering
    \begin{subfigure}[b]{0.49\textwidth}
        \includegraphics[width=7.5cm]{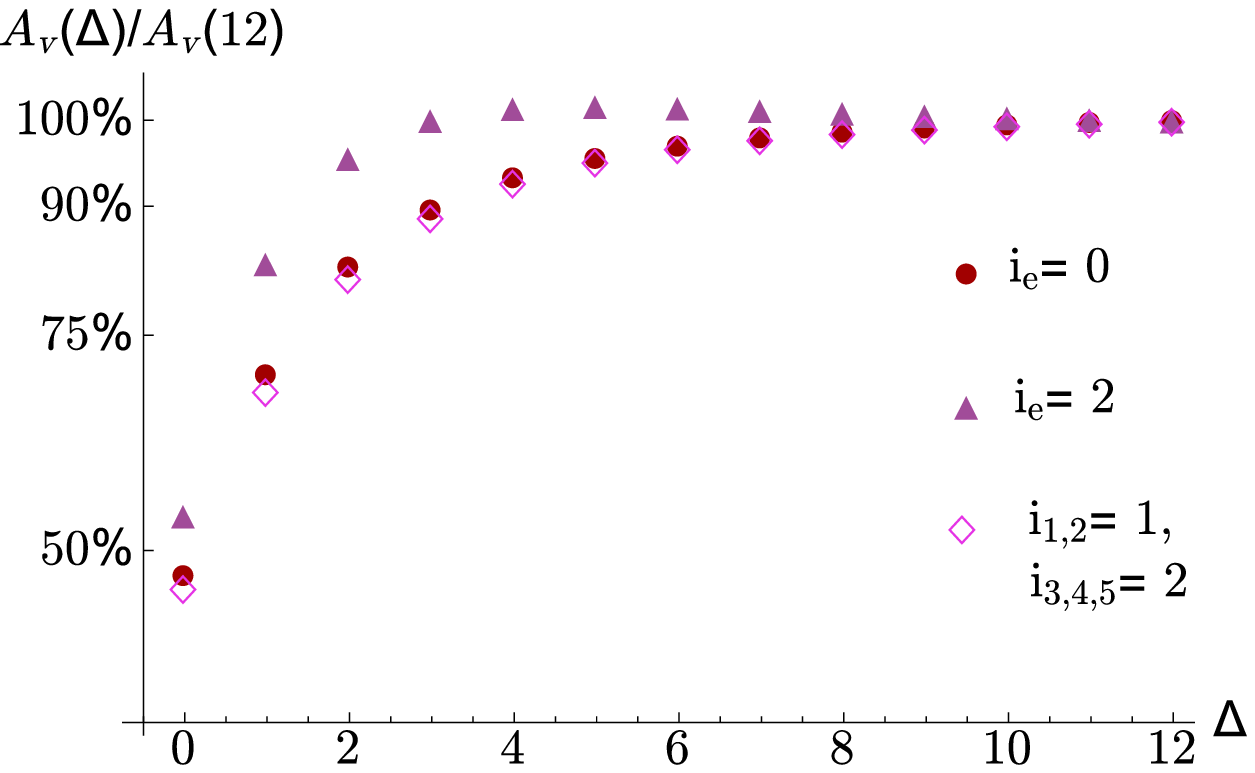}
    \end{subfigure}
    \begin{subfigure}[b]{0.49\textwidth}
        \includegraphics[width=7.5cm]{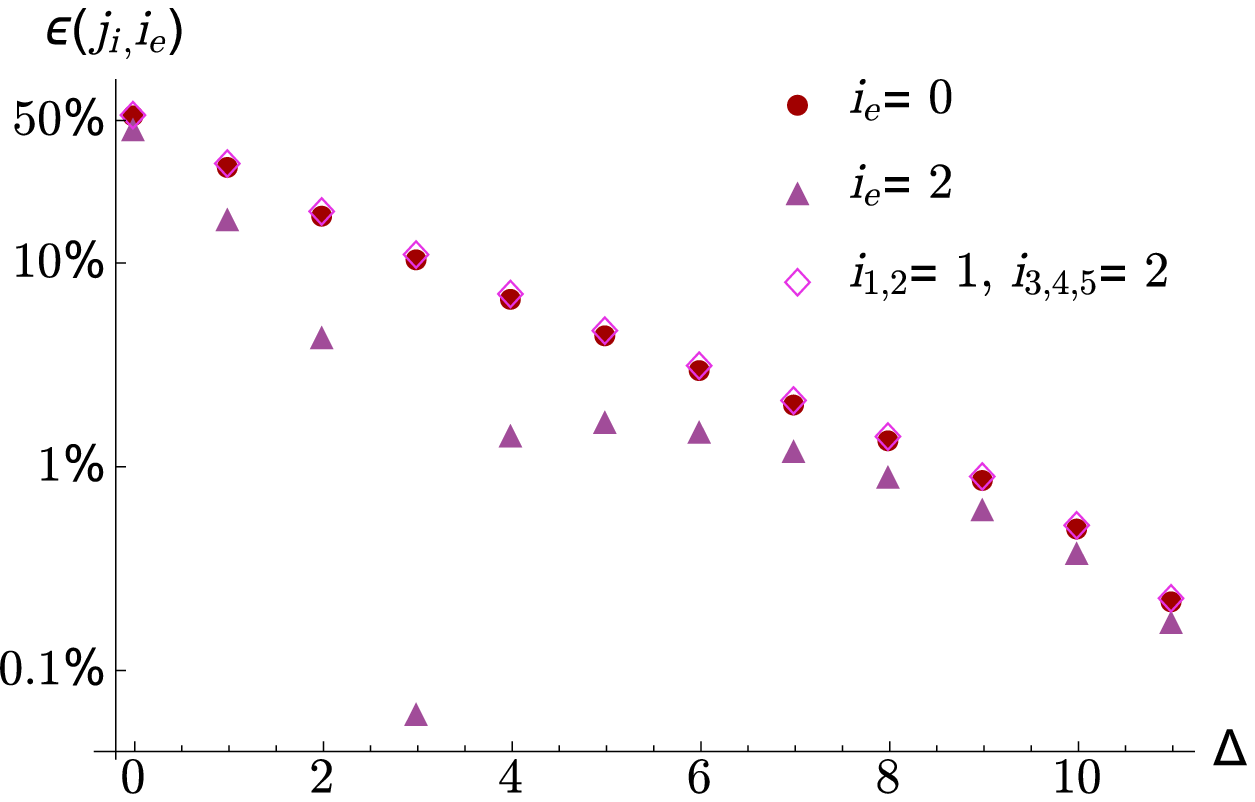}
    \end{subfigure}
      \caption{\label{fig:4SimplexConvergence}  \small{\emph{Convergence of the vertex amplitude.} Left panel: \emph{Convergence of the vertex amplitude while increasing the cutoff $\Delta$ for a configuration with all $j_f=1$ and three different set of boundary intertwiners. } Right panel: \emph{ The relative error $\epsilon$ between the shell $\Delta$ and the shell $\Delta=12$, that has been taken as the correct value of the amplitude. The error for the simplified model ($\Delta=0$) is $\sim 50\%$ while from $\Delta=5$ it decreases to $1\%-5\%$. The convergence of the vertex amplitude depends on many factors including the chosen boundary configuration and the Immirzi parameter.   } } }
\end{figure}

In the previous two sections we illustrated the evaluation of the constituents of the vertex amplitude \eqref{eq:vertexEPRL}. In this section we will discuss how all the contributions are summed together. The summation is performed over six auxiliary spins $l_f$ and four $k_e$ intertwiners, whose ranges depends on the $l_f$ but are generally bounded. The main problem in this step is the extension of these summations. They are, generally, unbounded and numerous.
As we already stressed multiple times, the full amplitude is well defined \cite{Engle:2008ev}, implying that the summations over the $l_f$ are convergent, thus being unbounded is not a problem. Nevertheless, to be able to do a numerical evaluation of the sum we need to introduce a homogeneous cut-off $\Delta$.  In the following, we will denote as the N\textsuperscript{th} \textit{shell} the set of factors contributing to the sum with maximum spin difference $\mathrm{Max}\left\lbrace l_f-j_f \right\rbrace = N$.
To be concrete we studied the convergence of the amplitude as a function of the cutoff, generally, it is quite rapid. We report some explicit examples in some particular cases in Figure \ref{fig:4SimplexConvergence}. 
It is not possible to have a unique prescription to set the optimal $\Delta$, since the convergence of the amplitude depends on the details of the data like the face spins $j_f$ and the Immirzi parameter. We explored other configurations and the convergence is qualitatively the same. However, the ratio between the amplitude value with a different number of shells varies.

\begin{figure}[H]
\begin{center}
\includegraphics[width=\textwidth]{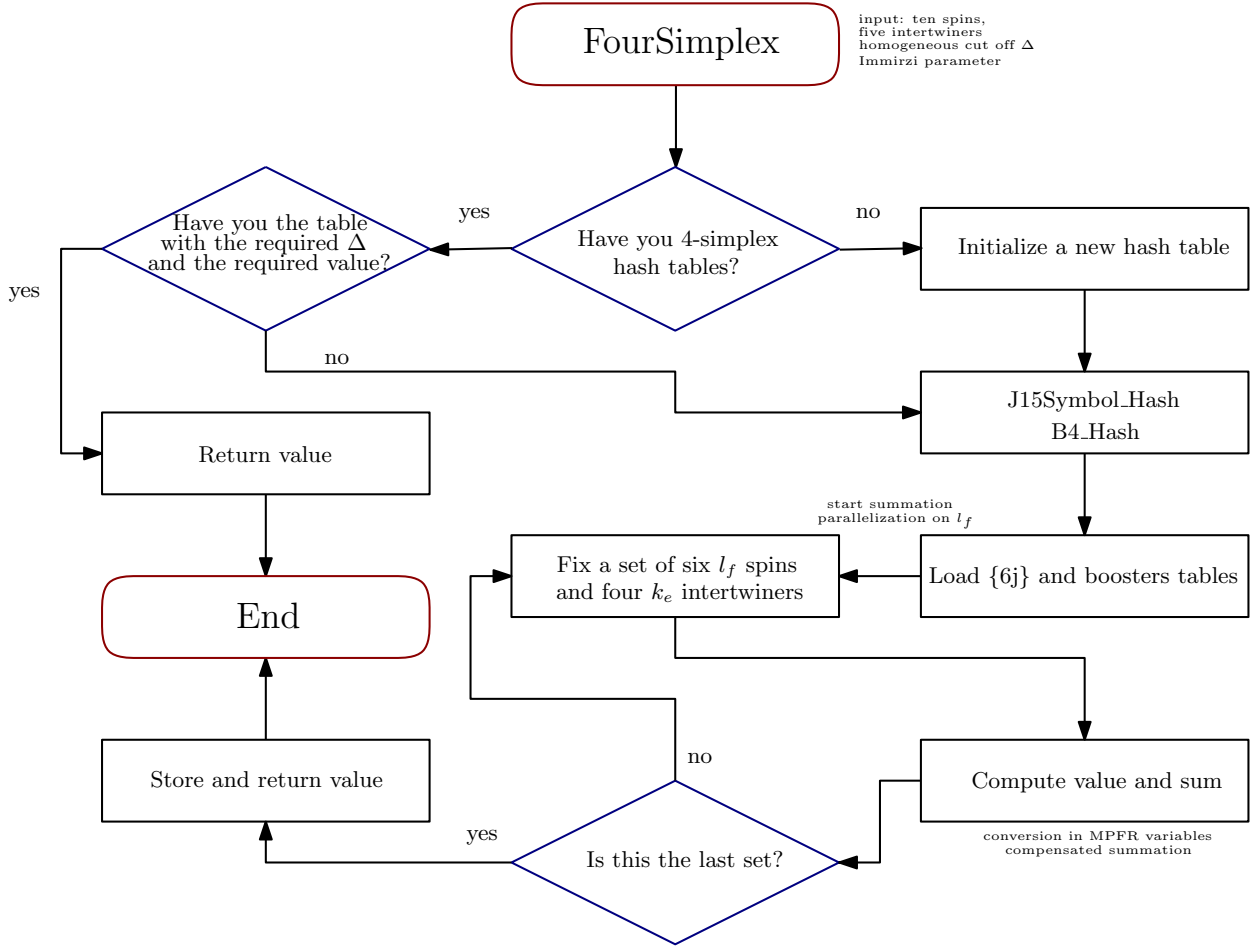}
\end{center}
\caption{\texttt{FourSimplex} is the main function to compute and store EPRL vertex amplitudes}
\label{FlowChart_FourSimplex} 
\end{figure}

In the library \coolname the function \texttt{FourSimplex} performs the summation of the 4-simplex vertex amplitude. The algorithm at the base of the implementation is described in Figure \ref{FlowChart_FourSimplex}. This function is the main result of this work and the big star of the library. Given the ten face spins, the five edge intertwiners, the cutoff $\Delta$ and the Immirzi parameter, \texttt{FourSimplex} uses the functions \texttt{J15Symbol\textunderscore Hash } and \texttt{B4\textunderscore Hash} described in the previous sections to precompute all the needed SU(2) invariants and booster functions and then proceed to the summations. 

\begin{figure}[H]
    \centering
        \includegraphics[width=7.5cm]{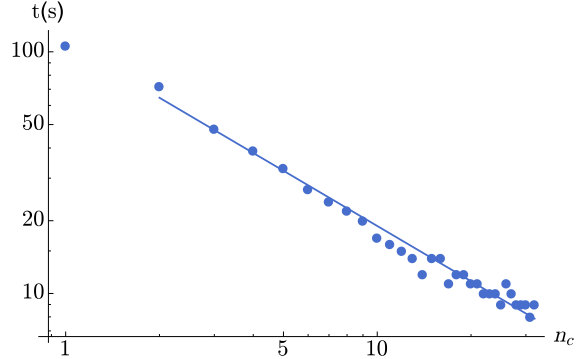}
      \caption{\label{fig:4SimplexCore}  \small{\emph{ Parallelization at fixed workload: Scaling of the time required by the computation of a 4-simplex configuration with $j_f=1$ and $\Delta = 11$ while increasing the number of cores. The power law $a\times j^b$ is given by $a=109s$ and $b=-0.76$. Since the parallelization acts only on the $l_f$ summations and not on the $k_e$ ones the code performs better, in this regard, when a large $\Delta$ is considered.  } } }
\end{figure}

This last step needs particular care. Given the vast number of factors to be summed over we need to employ a compensated summation algorithm to reduce the numerical error caused by the sum of finite precision floating point numbers. Furthermore, to minimize the error, we convert and multiply the components of the factors to arbitrary precision floating point numbers (using the library \texttt{MPFR}). Parallelization is needed at this step but, while we have implemented it for the summation over the face spins $l_f$, it was counterproductive to do it also for the four intertwiner sums $k_e$ because of triangular inequalities: when we fix a set of $l_f$ we set the bounds over the $k_e$ and the remaining four sums depend on them. In Figure \ref{fig:4SimplexCore} we show a parallelization test at fixed workload while increasing the number of cores.
 
The data are saved using the same convention as the boosters functions, we index them as \texttt{2j5.2j6.2j9.2j10\textunderscore 2$\Delta$}.
We stress that selection rules, to find the important terms within the sums, would be an essential step forward. At the present stage, we can set an empirical cut-off over boosters to reduce the number of terms to be summed over but it would be exciting to study, with new methods such as machine learning's techniques, selection rules for the vertex amplitude.

\begin{figure}[H]
    \centering
    \begin{subfigure}[b]{0.49\textwidth}
        \includegraphics[width=7.5cm]{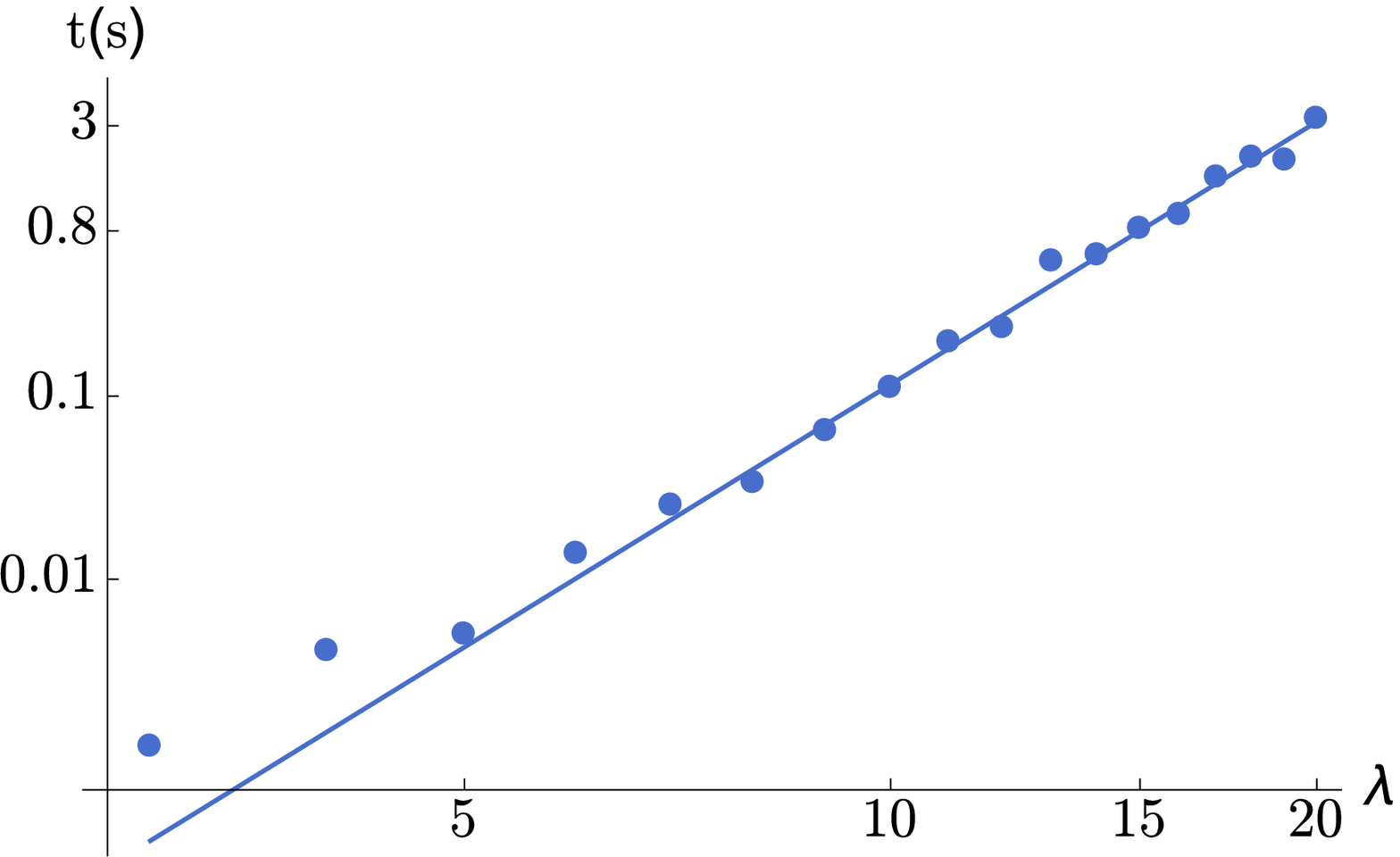}
    \end{subfigure}
    \begin{subfigure}[b]{0.49\textwidth}
        \includegraphics[width=7.5cm]{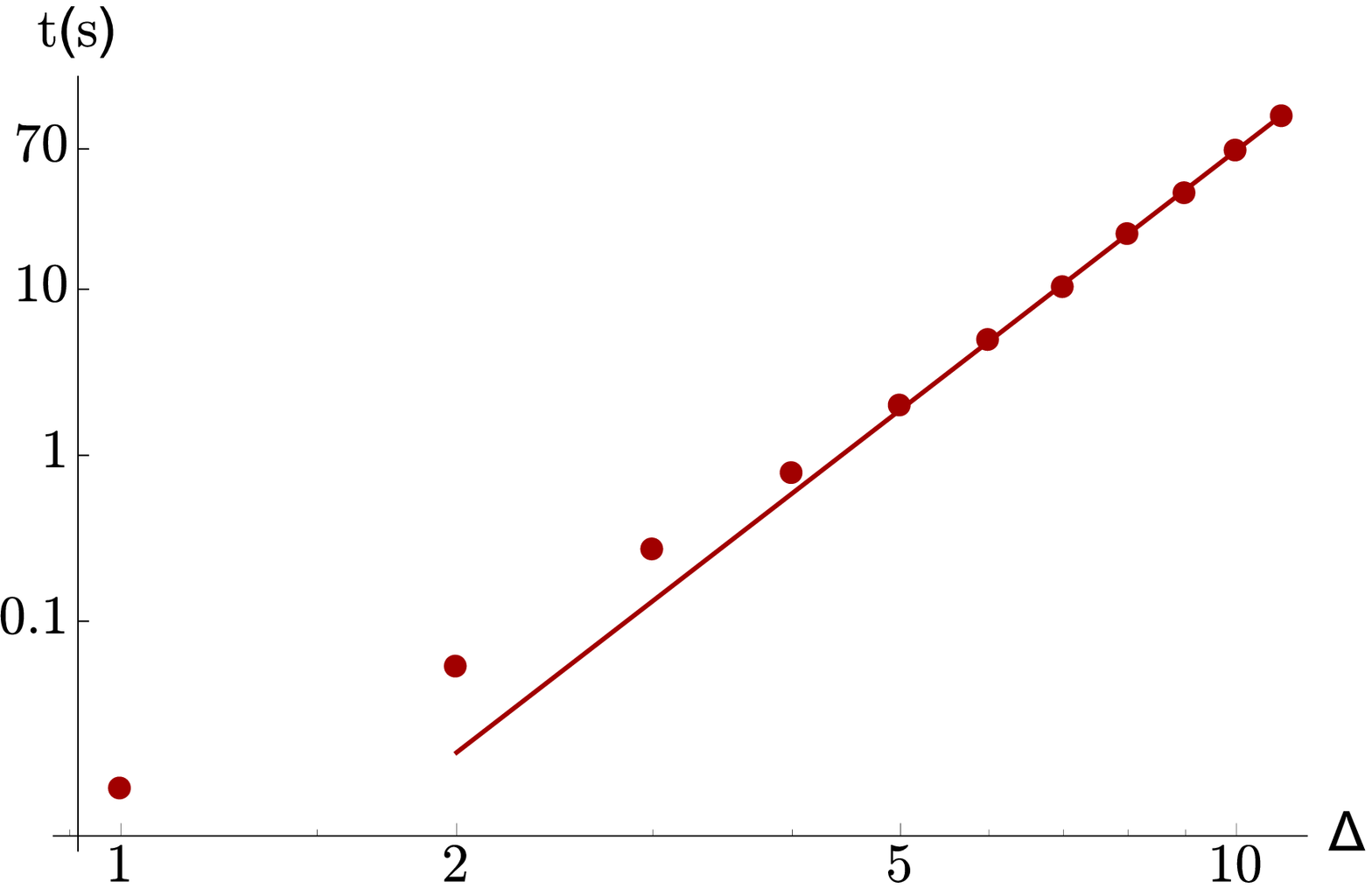}
    \end{subfigure}
    \caption{\label{fig:4SimplexTiming1}  \small{\emph{Evaluation time of the function} \texttt{FourSimplex}. Left panel: \emph{Scaling of the configuration with all the spins and the gauge fixed intertwiner equal $j_f=i_4=\lambda$ and $\Delta=0$. We see a power law trend $a\times j^b$ with $a=2.0\times 10^{-6}s$ and $b=4.8$. } Right panel: \emph{Scaling in the cutoff of the configuration with all the spins and the gauge fixed intertwiner equal to $j_f=i_4=1$. We see a power law trend $a\times j^b$ with $a= 4\times 10^{-4}s$ and $b= 5.2$.} } }
    \end{figure}

\begin{figure}[H]
    \centering
    \begin{subfigure}[b]{0.49\textwidth}
        \includegraphics[width=7.5cm]{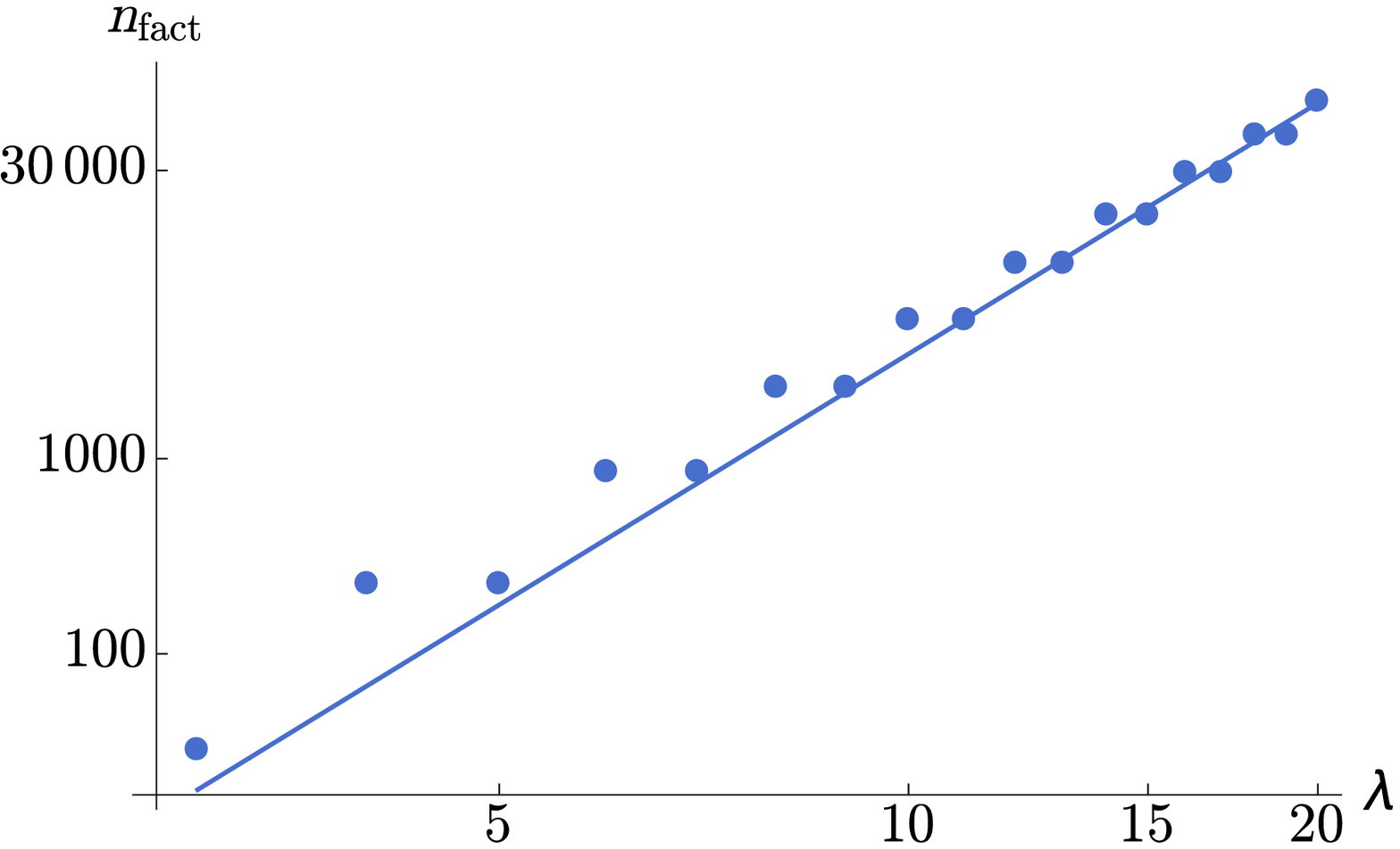}
    \end{subfigure}
    \begin{subfigure}[b]{0.49\textwidth}
        \includegraphics[width=7.5cm]{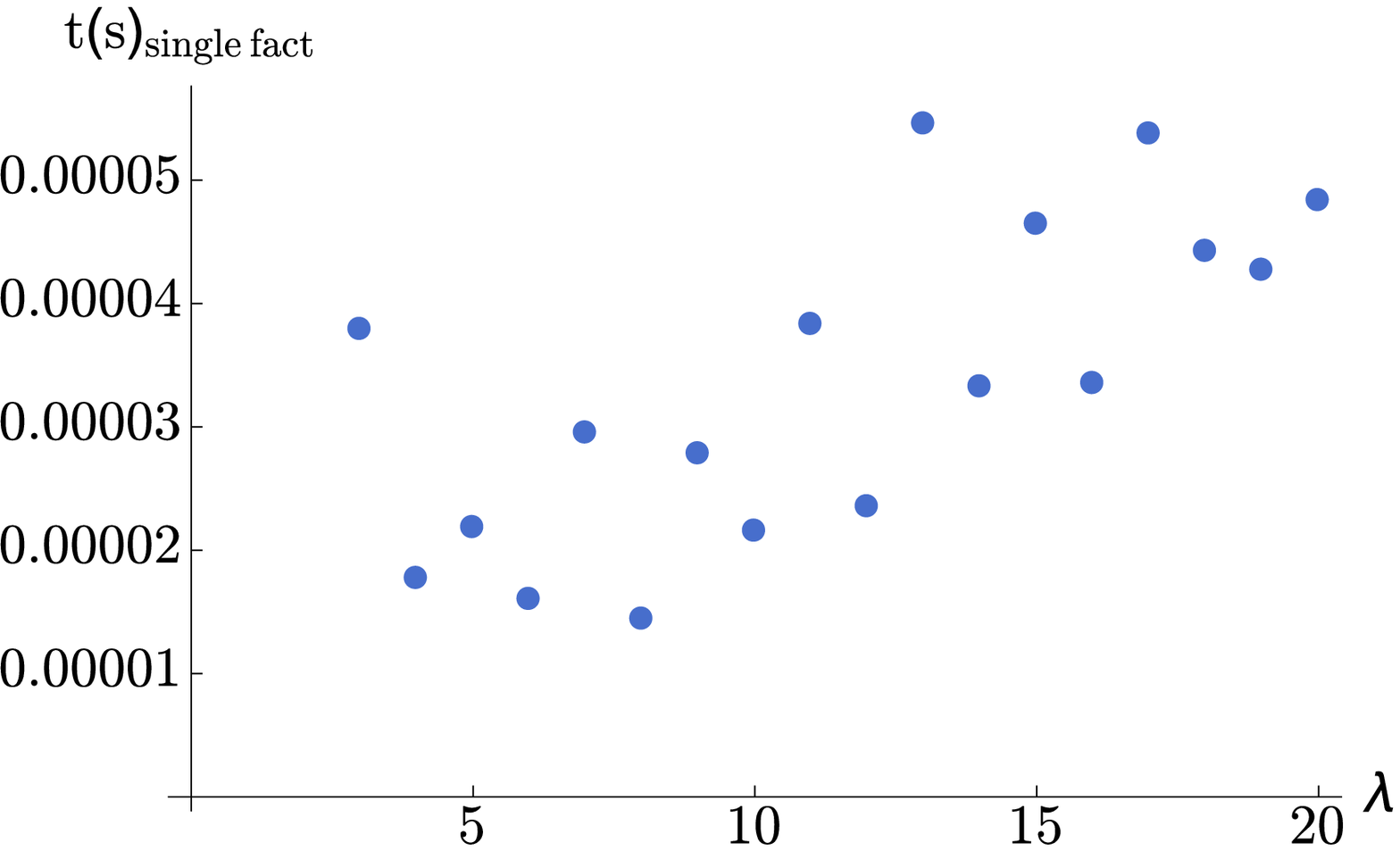}
    \end{subfigure}
      \caption{\label{fig:4SimplexTiming2}  \small{Left panel: \emph{We show the number of factors $\#_{fact}$ to be cycled over to compute a single 4-simplex with boundary spins, intertwiners equal to $\lambda$, and $\Delta  = 0$. The power law is given by  $a= 0.2$ and $b= 4.3$} Right panel: \emph{ Evaluation time of a single factor in a computation with all boundary spins and intertwiners equal to $\lambda$. Each factor takes $\sim10^{-5}$ s. Memory data access using \texttt{khash} works at $\sim10^{-7}$s: we have to retrieve data for four B4, two ``fixed'' $\{6j\}$ and for a sum of three $\{6j\}$, converting data to \texttt{MPFR} variables and then assembly the amplitude.This leads us to  $\sim10^{-5}$s of evaluation time for a single factor of the $l_f, k_e$ sum. If we would be able in the future to directly store $\{9j\}$ symbols we could save one order of magnitude in time. } } }
\end{figure}

\section{Utilities}
\subsection{Coherent states}
\label{sec:coherent}
The study of the semiclassical behavior of a spin foam transition amplitude relies on the relation between polyhedra and  SU(2) invariants \cite{Barrett:2009gg,Dona:2017dvf} and coherent intertwiners \cite{Livine:2007vk}. Thus, in \coolname, we provide a tool to compute Livine-Speziale coherent intertwiners as arbitrary precision floating point using \texttt{MPFR}. We refer to \cite{Dona:2017dvf} for the conventions and background material used in the present work, while we report here the definitions to fix the notation. We recall that an $n$-valent coherent intertwiner \cite{Livine:2007vk} is defined by the group averaging:
\begin{equation}
\label{eq:LivineSpeziale}
|\ket{\{j_a,\vec n_a\}} := \int dg \otimes_{a=1}^n g\ket{j_a,n_a} \ \in \ {\rm Inv}\otimes_{a=1}^n V^{(j_a)} \ ,
\end{equation}
where $V^{j_a}$ are SU(2) irreducible representation, $\ {\rm Inv}\otimes_{a=1}^n V^{(j_a)}$ is a singlet vector space and the vectors $n_a$ can be parametrized as $\vec {n_a}:=(\sin\Theta\cos\Phi, \sin\Theta\sin\Phi,\cos\Theta)$. The library provides a function to compute them for $n=4$. In this case the decomposition of a coherent intertwiner, in the recoupling channel $i$ with outgoing links, reads
\begin{equation}
\label{eq:coeffCS}
c_{i}(\vec n_a) := \bra{j_a,i}j_a, \vec n_a\ra = \sum_{m_a} \Wfour{j_1}{j_2}{j_3}{j_4}{m_1}{m_2}{m_3}{m_4}{i} \prod_{a=1}^4
D^{j_a}_{m_a,j_a}(\vec{n_a})
=  \raisebox{-5mm}{\includegraphics[width=2.8cm]{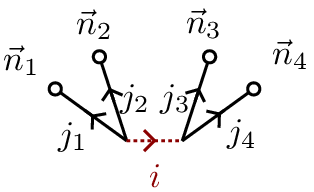}} \ ,
\end{equation}
where $D^{j}_{m,j}(\vec{n})=D^{j}_{m,j}(\phi, \Theta,-\phi)$ is the SU(2) Wigner matrix parametrized as:
\begin{equation}
\nn
D^{j}_{m,j}(\phi, \Theta,-\phi)= e^{-i m \phi} d^j_{m,j}(\Theta) e^{i j \phi} \ ,
\end{equation} 
with $d^j_{m,j}$ as a small Wigner matrix \cite{Varshalovich}.

\texttt{CoherentStateI} computes a coherent state coefficient $c_i (\vec n_a)$ for a given intertwiner $i$. It takes as input the four spins, one intertwiner, the four set of angles $(\Theta_a, \phi_a )$ (parametrizing the four normals $\vec{n}_a$) and also four signs related to the orientation of the links. 
For instance, to compute the coherent intertwiner in \eqref{eq:coeffCS} one needs to set:
 
\vspace{1em}
\texttt{CoherentStateI} ($2 i, 2 j_1, 2 j_2, 2 j_3, 2 j_4$, $\phi_1$, $\Theta_1$, $\phi_2$, $\Theta_2$, $\phi_3$, $\Theta_3$, $\phi_4$, $\Theta_4$, +1, +1, +1, +1) .
\vspace{1em}

\noindent The last four arguments represent the ``outgoing'' orientation of the four links in the same order the spins are inputted. To represent ``ingoing'' links is sufficient to flip the sign to $-1$ of the corresponding entries.

%----------------------------------------------------------------------------
\subsection{$SL(2,\mathbb{C})$ symbols}
\label{sec:sl2c}

In this section we review the basic formulas of Lorentzian recoupling theory and introduce the code of \coolname to compute them. We will follow closely the notation in \cite{Speziale:2016axj}. The $SL(2,\mathbb{C})$ Clebsch-Gordan coefficients can be factorized in an SU(2) one and a boost one:
\begin{equation}
\label{Cfactor}
C^{\rho_3 k_3 j_3 m_3}_{\rho_1k_1j_1m_1\rho_2k_2j_2m_2} = \chi(\rho_1,\rho_2,\rho_3,k_1,k_2,k_3;j_1,j_2,j_3) \, C^{j_3m_3}_{j_1m_1j_2m_2} \ .
\end{equation}
We will use a simplified notation indicating with $\chi(j_1,j_2,j_3)=\chi(\rho_1,\rho_2,\rho_3,k_1,k_2,k_3;j_1,j_2,j_3)$. Denoting with
\begin{equation}
J=\sum_i j_i, \qquad K=\sum_i k_i, \qquad P=\sum_i \rho_i,
\end{equation}
we can write the boost part of the $SL(2,\mathbb{C})$ Clebsch-Gordan coefficients as a finite sum of ratios of complex gamma functions as:
\begin{align}
\label{chiKer}
 \chi(j_1,j_2,j_3) &= \frac{(-1)^{\frac{K+J}2} }{4\sqrt{2\pi} } {\it x}(\rho_i,k_i) 
\Gamma(\tfrac{1-iP+K}{2}) \Gamma(\tfrac{1-iP-K}{2}) \sqrt{d_{j_1}d_{j_2}d_{j_3}} \, \kappa(\rho_i,k_i;j_i),
\end{align}
with
\begin{align}\label{kgen}
\kappa(\rho_i,k_i;j_i) & = (-1)^{ - k_1 - k_2} e^{-i\big(\Phi^{\rho_1}_{j_1}+\Phi^{\rho_2}_{j_2}-\Phi^{\rho_3}_{j_3}\big)}  \frac{(-1)^{j_1-j_2+j_3}}{\sqrt{d_{j_3}}}
\left[\frac{(j_1-k_1)!(j_2+k_2)!}{(j_1+k_1)!(j_2-k_2)!}\right]^{1/2}\\\nn & \times
\sum_{n=-j_1}^{{\rm min}\{j_1,k_3+j_2\}} \left[ \frac{(j_1-n)!(j_2+k_3-n)!}{(j_1+n)!(j_2-k_3+n)!} \right]^{1/2}
\, \Wthree{j_1}{j_2}{j_3}{n}{k_3-n}{-k_3}
\\\nn & \times 
\sum_{s_1={\rm max}\{k_1,n\}}^{j_1} \sum_{ s_2={\rm max}\{-k_2,n-k_3\}}^{j_2} \frac{(-1)^{s_1+s_2 -k_1+k_2}  \, (j_1+s_1)!(j_2+s_2)!}{(j_1-s_1)!(s_1-k_1)!(s_1-n)!(j_2-s_2)!(s_2+k_2)!(k_3-n+s_2)!} 
\\\nn & \times 
\frac{\Gamma(\tfrac{1-i (\rho_{1}- \rho_{2}- \rho_{3})-K+2 s_{1}}{2}) \Gamma(\tfrac{1+i (\rho_{1}- \rho_{2}+ \rho_{3})+K+2 s_{2}}{2}) 
\Gamma(\tfrac{1-i (\rho_{1}+\rho_{2}- \rho_{3})-k_{1}+k_{2}+k_{3}-2 n+2 s_{1}+2 s_{2}}{2})}{\Gamma (1-i \rho_{1}+s_{1}) \Gamma (1-i \rho_{2}+s_{2}) \Gamma (1+i \rho_{3}+s_{1}+s_{2}) \Gamma(\tfrac{1-i P 
-k_{1}+k_{2}+k_{3}-2 n}{2})},
\end{align}
where the additional phase 
\begin{align}\label{xdef}
{\it x}(\rho_i,k_i)  &= 
\frac{\Gamma\big(\tfrac{1+ i P - K)}{2}\big)}{|\Gamma\big(\tfrac{1+i P- K)}{2}\big)|}
\\\nn & \qquad \times \frac{\Gamma\big(\tfrac{1-i (-\rho_{1}+ \rho_{2}+ \rho_{3})-k_1+k_2+k_3}{2}\big)}{|\Gamma\big(\tfrac{1-i (-\rho_{1}+ \rho_{2}+ \rho_{3})-k_1+k_2+k_3}{2}\big)|} 
\frac{\Gamma\big(\tfrac{1-i (\rho_{1}- \rho_{2}+ \rho_{3})+k_{1}-k_{2}+k_{3}}{2} \big)}{|\Gamma\big(\tfrac{1-i (\rho_{1}- \rho_{2}+ \rho_{3})+k_{1}-k_{2}+k_{3}}{2} \big)|} 
\frac{ \Gamma\big(\tfrac{1-i (-\rho_{1} - \rho_{2} + \rho_{3})-k_{1}-k_{2}+k_{3}}{2}\big)}{| \Gamma\big(\tfrac{1- i (-\rho_{1} - \rho_{2} + \rho_{3}) -k_{1}-k_{2}+k_{3}}{2}\big)|},
\end{align}
was introduced in \cite{Speziale:2016axj} to make the Clebsch-Gordan coefficients, and consequently the booster functions, always real. This phase is different from the convention used in \cite{Ruhl:1970}.

We can write a $SL(2,\mathbb{C})$ version of the $(3jm)$ symbol. It reads
\begin{equation}\label{WignerSL2C}
\Wthree{\rho_1,k_1}{\rho_2,k_2}{\rho_3,k_3}{j_1}{j_2}{j_3} = (-1)^{j_1-j_2+j_3} \sqrt{d_{j_3}} \, \chi(j_1,j_2,j_3).
\end{equation}
We provide a function, \texttt{wigner\textunderscore 3j\textunderscore sl2c}, to compute it. Similarly we can also define an analog of the SU(2) $\{6j\}$ symbol for $SL(2,\mathbb{C})$ as:
\begin{align}
\{6\rho,6k\} = \sum_{j_i \geq k_i} \{6j_i\} &
\Wthree{\rho_1,k_1}{\rho_2,k_2}{\rho_3,k_3}{j_1}{j_2}{j_3}\Wthree{\rho_1,k_1}{\rho_5,k_5}{\rho_6,k_6}{j_1}{j_5}{j_6} \\\nn & \times
\Wthree{\rho_4,k_4}{\rho_2,k_2}{\rho_6,k_6}{j_4}{j_2}{j_6}\Wthree{\rho_4,k_4}{\rho_5,k_5}{\rho_3,k_3}{j_4}{j_5}{j_3},
\end{align}
To perform a numerical evaluation of this symbol we artifically cutoff the summation to $j_i\leq\Delta$. We provide \texttt{wigner\textunderscore 6j\textunderscore sl2c} to compute it. Since the $\kappa$ in \eqref{kgen} is given by a sum of ratio of complex gamma functions, an arbitrary precision computation is needed. We implemented it using \texttt{MPFR} variables. We tested the commands of this section up to spins $\approx 50$, we leave the improvement of this upper limit to future works.

\section{What comes next?}
In this paper, we described the library \texttt{sl2cfoam} that we developed to compute the spin foam vertex amplitude defined by the EPRL model.

Many technical improvements are necessary to unlock the full potential of this code. In analogy to what we do for the $\{6j\}$ symbols, we need to implement hash tables also for $\{9j\}$ symbols. This would speed up the computation of almost an order of magnitude, trading at low spins memory usage with faster access to data. 
Moreover, we want to explore the possibility of computing booster functions with the Kerimov formula, as explained in \cite{Speziale:2016axj}, for which recursion relations are available, simplifying the summation over the auxiliary spins $l_f$. 
To perform the computation of transition amplitudes with many vertices, it is crucial to improve our parallelization process, using \texttt{MPI} and \texttt{POSIX} threads enabling many servers calculations. A similar improvement will be reached by finding selection rules to restrict the summation to only the important terms. Our lack of analytical insight into the booster functions will force us to use new techniques like for example machine learning or Monte Carlo techniques. 
Finally, we want to search for a dynamical way of dealing with the cutoff $\Delta$ to be able to perform computations with an error set a priori.
We plan to explore these fascinating possibilities soon.

\noindent For a quick reference, we report in Table \ref{table:timings} the timing of the various functions in \coolname. 
\begin{table}
\begin{center}
\bgroup
\def\arraystretch{1.5}%  1 is the default, change whatever you need
\begin{tabular}{c|c|c|c|c|c|c|c|c|}
 & \multicolumn{2}{c|}{$j_{f}=i_{4}=7$} & \multicolumn{2}{c|}{$j_{f}=i_{4}=15$} & \multicolumn{2}{c|}{$j_{f}=1,$ $i_{4}=1,$ $\Delta=5$} & \multicolumn{2}{c|}{$j_{f}=1,$ $i_{4}=1,$ $\Delta=10$}\tabularnewline
\hline 
\multirow{2}{*}{\hspace{0.1cm} ${15j}$ \hspace{0.1cm}} & $0.1s$ & $6.8\mathtt{Mb}$ & $6.1s$ & $218\mathtt{Mb}$ & $1.4s$ & $0.8\mathtt{Mb}$ & $62.0s$ & $6.8\mathtt{Mb}$\tabularnewline
\cline{2-9} 
 & \multicolumn{2}{c|}{$a=2.9\cdot10^{-6}s$} & \multicolumn{2}{c|}{$b=5.4$} & \multicolumn{2}{c|}{$a=4\cdot10^{-4}s$} & \multicolumn{2}{c|}{$b=5.0$}\tabularnewline
\hline 
\multirow{2}{*}{\hspace{0.1cm}$B_{4}$\hspace{0.1cm}} & \multicolumn{2}{c|}{$99.0s$} & \multicolumn{2}{c|}{$642.1s$} & \multicolumn{2}{c|}{$2433s$} & \multicolumn{2}{c|}{$49214s$}\tabularnewline
\cline{2-9} 
 & \multicolumn{2}{c|}{$a=0.7s$} & \multicolumn{2}{c|}{$b=2.5$} & \multicolumn{2}{c|}{$a=2.7s$} & \multicolumn{2}{c|}{$b=4.3$}\tabularnewline
\hline 
\multirow{2}{*}{\hspace{0.1cm}$A_{v}$\hspace{0.1cm}} & \multicolumn{2}{c|}{$0.03s$} & \multicolumn{2}{c|}{$0.08s$} & \multicolumn{2}{c|}{$2.0s$} & \multicolumn{2}{c|}{$69.7s$}\tabularnewline
\cline{2-9} 
 & \multicolumn{2}{c|}{$a=2.0\cdot10^{-6}s$} & \multicolumn{2}{c|}{$b=4.8$} & \multicolumn{2}{c|}{$a=4\cdot10^{-4}s$} & \multicolumn{2}{c|}{$b=5.2$}\tabularnewline
\hline 
\end{tabular}
\egroup
 \caption{\label{table:timings}  \small{\emph{Summary of the computational time of the functions }
 \texttt{J15Symbol\textunderscore Hash}, \texttt{B4\textunderscore Hash}, \emph{and} \texttt{FourSimplex}\emph{. We report some examples of computational time for configuration with all spins equal to $7$ and $15$ in the first two columns and the parameters of the power law behavior $a j^b$. In the last two columns we report some examples of computational time for configuration with all spins equal to $1$ as a function of the cutoff and the parameters of the power law behavior $a \Delta^b$. All the computations are performed with a single processor. }}}
\end{center}
\end{table}

We firmly believe that this code could profoundly change the way we understand the dynamics of loop quantum gravity. One of the most successful computations in the theory proved the connection in the semiclassical limit with discrete General Relativity \cite{Barrett:2009gg,Barrett:2011xa}. They showed that a single EPRL vertex amplitude contains the exponential of Regge action for a Lorentzian 4-simplex in the large spin limit. While this calculation is of dramatic importance for the theory, its reign of validity is limited to a single simplex. We still lack a definite result for an extended triangulation made of many simplices glued together. This limitation made the community questions if the simplicity constraints, responsible of constraining the original topological BF theory to General Relativity, are enforced correctly\cite{Conrady:2008mk, Bonzom:2009hw, Hellmann:2013gva}. This translate to an inability of the model in describing amplitudes compatible with curved spacetime. Preliminary results, limited to the Euclidean version of the model, show that curved geometries are indeed allowed in the asymptotic limit of the amplitude. Using \coolname we will investigate the large spin limit of the $\Delta_3$ complex, dual to three 4-simplices sharing a bulk triangle, the most straightforward triangulation with an internal face.

We are performing a consistency check of our code by computing the homogeneous scaling of the Lorentzian single vertex amplitude with Euclidean boundary data. To do so, we calculated the vertex amplitude with spins, both integers and half-integers up to order $10$, contracted with coherent intertwiner states representing the boundary of a regular Euclidean four simplex. We sum two shells per amplitude with an estimated error of approximately the $20\%$. We find an amazing agreement with the analytic formula of \cite{Barrett:2009gg}. 

We also plan to explore numerically the divergence of the theory recently estimated in \cite{Riello:2013bzw, Dona:2018pxq} by computing the ``self-energy'' diagram. 
A large set of power-law divergences is still compatible with both estimates, and the tensorial structure of the diagram is still unknown. To confirm the first and compute the second will allow us to understand how to correctly regularize the theory without introducing a cosmological constant following a prescription similar to the one studied in \cite{Freidel:2002dw}.

The renormalization group flow of the EPRL vertex amplitude is of crucial importance for the future of the theory. Finding a fixed point in the flow and identifying phase transitions would allow us to understand if the diffeomorphism invariance, broken by the discretization, is recovered and what mechanism will restore it. An interesting insight on this topic is provided in \cite{Dittrich:2014ala, Dittrich:2016tys, Delcamp:2016dqo} and in \cite{Bahr:2014qza, Bahr:2016hwc, Bahr:2017klw, Bahr:2017eyi,Bahr:2018gwf}. In this last series of works they perform a numerical investigation of the flow of the Euclidean EPRL model and on a hypercubic lattice, with a large number of vertices. The calculations are possible thanks to some strong but vital approximations: they work with the Euclidean version of the theory, they restrict to a specific form of the intertwiners, and instead of using the full vertex amplitude they replace it with its asymptotic formula. At the present \coolname cannot directly attack analog computation, the number of vertices needed for a lattice computation is still out of our reach, but we do not exclude to be able to do it in a near future after all the proposed upgrade are implemented and we have access to more processing power (for the moment we are limited to a 32 cores machine).

Nevertheless, if we restrict to a purely quantum region characterized with relatively small spins, probably up to order $10$, we can now compute any transition amplitude with a few vertices. We would like to invite the community to look for interesting computations that could be performed within those limitations, and to develop a common framework in which the theory can be used to answer physically meaningful questions.

%--------------------------------------------------------------------------------------------------
\section*{Acknowledgments}
This work was supported in part by the NSF grants PHY-1505411, PHY-1806356 and the Eberly research funds of Penn State. We want to thank Fran\c{c}ois Collet and Francesco Gozzini for providing a basic code for the booster function and the complex gamma function with arbitrary precision floating points numbers and especially for many interesting and helpful discussions. Thanks to NORDITA and to the organizers of the workshop ``Quantum Gravity on the Computer'', a very inspiring and useful week in Stockholm. Many thanks to Simone %Spicy
Speziale for fruitful discussions and encouragement and we are very happy to welcome Emilio in this puzzling world.

%----------------------------------------------------------------------------
\appendix
%----------------------------------------------------------------------------

%----------------------------------------------------------------------------
\section{SU(2) Symbols}\label{app:su2symbols}
%----------------------------------------------------------------------------

We use the definition of the Wigner's $(3jm)$ symbol reported in \cite{Varshalovich} with the following orthogonality properties
\begin{equation}
\sum_{m_1,m_2} \Wthree{ j_1}{ j_2}{ j_3} {m_1}{m_2}{m_3} \Wthree{ j_1}{ j_2}{ j_3} {m_1}{m_2}{n_3} = \frac{\delta_{j_{3}l_{3}}\delta_{m_3n_3}}{2 j_3 + 1} \ ,
\end{equation}
implying they are normalized to one and that, if triangular inequalities are not satisfied, they vanish. We can define the $(4jm)$ symbol in the following way, as the contraction of two $(3jm)$ symbol using an intertwiner $i$
\begin{equation}
\Wfour{ j_1}{ j_2}{ j_3} {j_4}{m_1}{m_2}{m_3}{m_4}{i}\equiv \sum_{m_i} (-1)^{i-m_i} \Wthree{ j_1}{ j_2}{ i} {m_1}{m_2}{m_i} \Wthree{ i}{ j_3}{ j_4} {-m_i}{m_3}{m_4}\ ,
\end{equation}
with the orthogonality relations
\begin{equation}
\sum_{m_1,m_2,m_3} \Wfour{ j_1}{ j_2}{ j_3} {j_4}{m_1}{m_2}{m_3}{m_4}{i_1} \Wfour{ j_1}{ j_2}{ j_3} {j_4}{m_1}{m_2}{m_3}{m_4}{i_2} = \frac{\delta_{i_1 i_2}}{2 i_1 + 1} \frac{\delta_{j_{4}l_{4}}\delta_{m_4n_4}}{2 j_4 + 1} \ ,
\end{equation}
where $ \frac{\delta_{i_1 i_2}}{2 i_1 + 1}$ is the normalization factor.

\end{document}